\documentclass[
 reprint,
 amsmath,
 pre,
 amssymb,
 nofootinbib,
 aps,
 longbibliography,
]{revtex4-2}

\usepackage[pdftex]{graphicx}
\usepackage{amsfonts}
\usepackage{ifsym}
\usepackage{pifont}
\usepackage{footmisc}
\usepackage{colonequals}
\makeatletter
\newlength{\apb@width}
\newcommand{\autoparbox}[2][c]{\settowidth{\apb@width}{#2}\parbox[#1]{\apb@width}{#2}}

\makeatother

\newcommand{\namedref}[2]{\hyperref[#2]{#1~\ref*{#2}}}

\renewcommand{\Re}{\mathop{\mathrm{Re}}}
\renewcommand{\Im}{\mathop{\mathrm{Im}}}

\newcommand{\Csphere}{{}^\bullet\kern-1.2pt C}
\newcommand{\Ctorus}{{}^\circ\kern-1.2pt C}

\newcommand{\COMMENT}[1]{}

\newcommand{\neqa}{\nonumber\end{eqnarray}}
\newcommand{\la}[1]{\label{#1}}

\newcommand{\<}{{\langle}}
\renewcommand{\>}{{\rangle}}

\newcommand{\re}{\relax{\rm I\kern-.18em R}}

\def\su2{{SU(2)}}

\def\[{\left[}
\def\]{\right]}

\def\({\left(}
\def\){\right)}
\def\[{\left[}
\def\]{\right]}

\def\<{\langle}
\def\>{\rangle}

\def\i2{\frac{i}{2}}

\def\cN{{\cal N}}

\def\cP{{\cal P}}

\def\2F1{\,_2{\rm F}_1}
\def\sumint{\sum\hspace{-1.4em}\int}

\usepackage[usenames,dvipsnames]{xcolor}
\usepackage{color}
\usepackage{graphicx}
\usepackage{dcolumn}
\usepackage{bm}
\RequirePackage[colorlinks=true
,urlcolor=blue
,anchorcolor=blue
,citecolor=blue
,filecolor=blue
,linkcolor=blue
,menucolor=blue
,linktocpage=true
,pdfproducer=medialab
,pdfa=true
]{hyperref}

\usepackage{cancel}
\usepackage{ctable}
\usepackage{booktabs}
\newcolumntype{L}[1]{>{\raggedright\let\newline\\\arraybackslash\hspace{0pt}}m{#1}}
\newcolumntype{C}[1]{>{\centering\let\newline\\\arraybackslash\hspace{0pt}}m{#1}}
\newcolumntype{R}[1]{>{\raggedleft\let\newline\\\arraybackslash\hspace{0pt}}m{#1}}

\newcommand{\beq}{\begin{equation}}
\newcommand{\eeq}{\end{equation}}
\newcommand{\beqq}{\begin{equation*}}
\newcommand{\eeqq}{\end{equation*}}
\newcommand\beqa{\begin{eqnarray}}
\newcommand\eeqa{\end{eqnarray}}
\newcommand\beqaa{\begin{eqnarray*}}
\newcommand\eeqaa{\end{eqnarray*}}
\newcommand\bea{\begin{array}}
\newcommand\eea{\end{array}}

\begin{document}

\title{Constraining Glueball Couplings}

\author{Andrea L. Guerrieri$^{a,b,c}$, Aditya Hebbar$^d$ and Balt C.~van Rees$^d$}

\affiliation{
	$^a$Dipartimento di Fisica e Astronomia, Universita degli Studi di Padova, Italy \\ $^b$Istituto Nazionale di Fisica Nucleare, Sezione di Padova, via Marzolo 8, 35131 Padova, Italy \\
	$^c$Perimeter Institute for Theoretical Physics, Waterloo, Ontario N2L 2Y5, Canada }
\affiliation{$^d$CPHT, CNRS, École Polytechnique, Institut Polytechnique de Paris, 91120 Palaiseau, France}

\begin{abstract}
We set up a numerical S-matrix bootstrap problem to rigorously constrain bound state couplings given by the residues of poles in elastic amplitudes. We extract upper bounds on these couplings that follow purely from unitarity, crossing symmetry, and the Roy equations within their proven domain of validity. First we consider amplitudes with a single spin 0 or spin 2 bound state, both with or without a self-coupling. Subsequently we investigate amplitudes with the spectrum of bound states corresponding to the estimated glueball masses of pure $SU(3)$ Yang-Mills. In the latter case the `glue-hedron', the space of allowed couplings, provides a first-principles constraint for future lattice estimates.
\end{abstract}

\pacs{Valid PACS appear here}
\maketitle

\section{Introduction}
Glueballs are the stable, massive and colorless particles that appear in the spectrum of Yang-Mills theories at long distances. Describing their physics is extremely challenging. In the real world all glueballs are unstable, and isolating them as resonances is very difficult \cite{Crede:2008vw} because they carry the same quantum numbers as neutral mesons. On the lattice their spectrum has recently been measured in the $SU(N_c)$ pure Yang-Mills theories \cite{Athenodorou:2020ani,Athenodorou:2021qvs}. Determining their interactions is however substantially more difficult, but see~\cite{DEFORCRAND1985107,Yamanaka:2019yek,Giacosa:2021brl} for some attempts.

In this letter we use the S-matrix bootstrap to constrain three-point couplings between glueballs with a spectrum as in table \ref{SpectrumSU3} which should correspond to $SU(3)$ Yang-Mills theory. To demonstrate the general applicability of our method we will also constrain three-point couplings in other processes.

\begin{table}[h]
\centering
\begin{tabular}{||c | c  | c||} 
 \hline
  \quad & \quad $J^{PC}$ \quad & Mass  \\ [0.5ex] 
 \hline 
 \hline
\quad  G \quad & \quad $0^{++}$ \quad & 1\\
 \hline
 \quad  H \quad & \quad $2^{++}$ \quad & $1.437 \pm 0.006$ \\
 \hline
\quad  $G^*$ \quad & \quad $0^{++}$ \quad & $1.72\pm 0.01	$ \\
 \hline
 \quad  $H^*$ \quad & \quad $2^{++}$ \quad & $1.99\pm 0.01$ \\
 \hline
\end{tabular}
\caption{Lattice estimates for the spectrum of $P,C=+,+$ stable Glueballs \cite{Athenodorou:2020ani,Athenodorou:2021qvs}, in units where the lightest glueball has unit mass. The mass of the excited spin-two Glueball is very close to the two particle threshold.}
\label{SpectrumSU3}
\end{table}

Our approach will be to extend the `dual' S-matrix bootstrap \cite{Lopez:1976zs,Guerrieri:2021tak}, first to general elastic 2-to-2 amplitudes with bound state poles and then to the $GG \to GG$ scattering amplitude in particular. This leads to rigorous bounds that follow purely from proven analyticity, crossing symmetry and unitarity. This approach should be contrasted with the `primal' method of \cite{Paper2,Paper3} where approximate extremal amplitudes are constructed numerically and which we expect to apply to this problem in the near future.

\begin{figure}[h]
    \centering
    \includegraphics[scale=0.23]{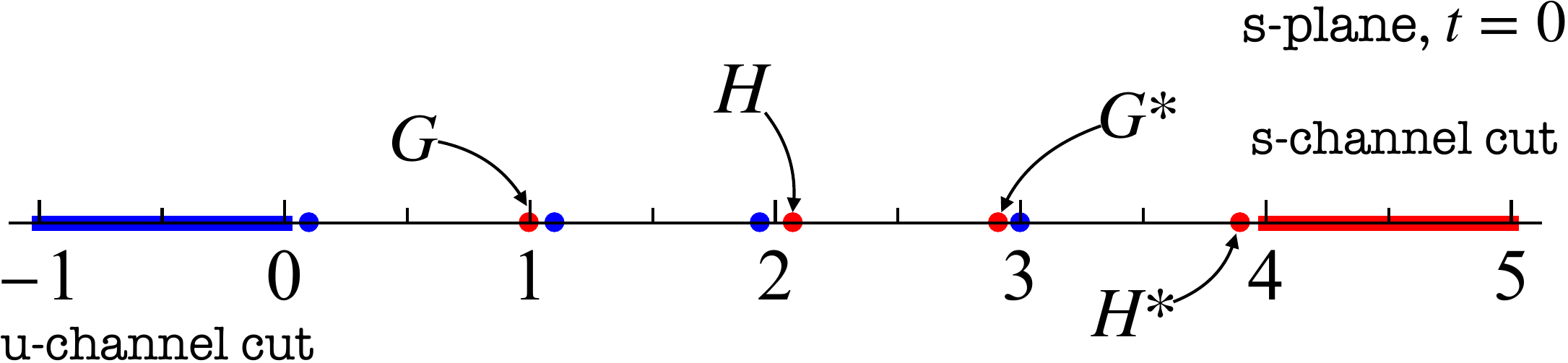}
    \caption{Singularity structure of the $GG\to GG $ amplitude in the $s$-plane at fixed $t=0$. In red, we denote the $s$-channel poles and cut, in blue the corresponding crossed $u$-channel singularities.}
    \label{fig:poles}
\end{figure}

To orient the reader we show part of the analytic structure of the forward $GG \to GG$ amplitude in figure \ref{fig:poles}. Shown are the normal thresholds as well as 8 simple poles associated with the stable glueballs of table \ref{SpectrumSU3}. The residues of the poles corresponding to particle $X$ are proportional to the squared three-point couplings $g_X^2$. It is these couplings that we will bound, with the allowed region in the three-dimensional space spanned by $(g_G, g_H, g_{G^*})$ shown in figure \ref{fig:glue-hedron} on page \pageref{fig:glue-hedron}. Before discussing these results in detail we will first explain our method.

\section{Scattering amplitudes with bound states poles}
In this section we review the dual $S$-matrix Bootstrap technology based on fixed-$t$ dispersion relations. It was developed in a sequence of papers during the 1970s \cite{Lopez:1974cq,Lopez:1975wf,Lopez:1975ca,Bonnier:1975jz,Lopez:1976zs} and more recently revisited and extended in \cite{Guerrieri:2021tak,Miro:2023bon}. Consider an amplitude $T(s,t)$ describing the $2\to 2$ scattering of the lightest scalar particles in the theory, with mass $m = 1$. Suppose the amplitude has several poles below threshold $p \in \cP$ associated to stable particles of different spins. Using the boundedness of the amplitude at fixed $t$ ~\cite{PhysRev.123.1053, Jin:1964zza, Martin:1965jj}
\beq
\label{Tasymptoticbehavior}
\lim_{s\to \infty}\frac{T(s,t< t^*)}{|s|^2}=0,
\eeq
where $t^*$ is the squared mass of the lightest particle with spin $\ell\geq 2$, we write the doubly-subtracted dispersion relation:
\begin{multline}
\label{bestdispersion}
  T(s,t) = T(s_0,t_0)+ \sum_{p \in P} g^2_p R_{\mu_p\, \ell_p}(s,t;s_0,t_0) \,  \\
  + \sumint\limits_{\ell,v}\, \Im f_\ell(v)  R_{v\, \ell}(s,t;s_0,t_0) 
\end{multline}
where the kernel $R_{v\, \ell} (s,t;s_0,t_0)$ is defined in appendix \ref{app:dispersionderivation}, which also contains a derivation of the above formula.

We project equation \eqref{bestdispersion} onto even spin $J$ partial waves by using:
\begin{equation}
  f_J(s) = \frac{\cN_d}{2} \int\limits_{-1}^1 dz (1-z^2)^{\frac{d-4}{2}}P_J^{(d)}(z) T\left(s,t_s(z)\right)
\end{equation}
with $\mathcal N_4 = (16 \pi)^{-1}$ and
\begin{equation}
t_{s}(z) = \frac{1}{2} (s-4) (1-z)\,.  
\end{equation}
This produces the Roy equations~\cite{Roy:1971tc}:
\begin{multline}
\label{royequations}
\Re f_J(s)  = \frac{\delta_{J,0}}{n_0^{(d)}}T(s_0,t_0) + \sum_{p \in \cP}g_p^2 R_{\mu_p\, \ell_p}^{(J)}(s;s_0,t_0)\\
+ \sumint\limits_{\ell,v} \, \Im f_\ell(v) R_{v\, \ell}^{(J)}(s;s_0,t_0) dv,  
\end{multline}
where we used that $P_0^{(d)} = 1$ and defined the spin $J$ projected kernel as
\begin{multline}
\label{kernelprojection}
R_{v\, \ell}^{(J)}(s;s_0,t_0) \colonequals \\
\frac{2 \cN_d}{2} \int\limits_0^1 dz (1-z^2)^{\frac{d-4}{2}}P_J^{(d)}(z) R_{v\, \ell}(s,t_s(z);s_0,t_0) \,.
\end{multline}
We can restrict the integration range from $0$ to $1$ because we take $J$ to be even. For the odd $J$ Roy equations we would obtain `null constraints' corresponding to the $t \leftrightarrow u$ crossing symmetry. We will however replace them with the alternative constraints obtained by demanding the vanishing of suitable derivative combinations around the crossing symmetric line given by $s = 4 -2 t$. For any point $t_c$ along this line we have:
\begin{equation}
  n \text{ odd} : \quad \left(\frac{\partial}{\partial \tau}\right)^n \left. T\left(4 - 2 t_c, t_c + \tau\right)\right|_{\tau = 0}  = 0
\end{equation}
and we can therefore write
\begin{multline}
\label{crossingconstraint}
  n \text{ odd} : 0 = \sum_{p \in P} g^2_p R^{[n]}_{\mu_p\, \ell_p}(t_c; s_0,t_0) \,  \\
  + \sumint\limits_{\ell,v}\, \Im f_\ell(v)  R^{[n]}_{v\, \ell}(t_c; s_0,t_0) 
\end{multline}
where 
\begin{multline}
  R^{[n]}_{v\, \ell}(t_c; s_0,t_0) \colonequals \partial_\tau^n \left.
  R_{v\, \ell}\left(4 - 2 t_c, t_c + \tau; s_0,t_0\right)
  \right|_{\tau = 0} 
\end{multline}
is the $n$'th transverse derivative of the kernel at $t_c$.

The unitarity constraint finally reads
\begin{equation}
  1 \geq | S_\ell(s) | = |1 + i \tilde \rho (s) f_\ell(s)|
\end{equation}
with $\tilde \rho_s = \sqrt{s-4}/\sqrt{s}$. It is equivalent to
\begin{equation}
\label{unitarityphysical}
  \begin{pmatrix}
  1 + \Re[S_\ell] & \Im[S_\ell]\\
  \Im[S_\ell] & 1 - \Re[S_\ell]
  \end{pmatrix} = 
  \begin{pmatrix}
  2 - \tilde \rho_s  \Im[f_\ell] & \tilde \rho_s  \Re[f_\ell]\\
  \tilde \rho_s \Re[f_\ell] &  \tilde \rho_s  \Im[f_\ell]
  \end{pmatrix}
  \succeq 0
\end{equation}
for all even $\ell$ and all $s \geq 4 m^2$. Below physical threshold the extended unitarity constraint is just:
\begin{equation}
\label{unitarityextended}
  g_p^2 \geq 0\,.
\end{equation}

\section{Duality}
Our \emph{primal variables} are (the real and imaginary parts of) $f_\ell(s)$ for all even $\ell$ and all $s \geq 4$, the squared couplings $g_p^2$, and the subtraction constant $T(s_0,t_0)$. They are subject to the linear constraints \eqref{royequations}, \eqref{crossingconstraint} and the unitarity inequalities \eqref{unitarityphysical} and \eqref{unitarityextended}. Within this space we aim to find maximal allowed values of (some linear combination of) the squared couplings, an objective which we formulate as 
\begin{equation}
  \max \sum_{p} v_p  g_p^2
\end{equation}
for some fixed vector $v_p$. This is immediately seen to define a continuum-version of a semidefinite program, and in this section we formulate its dual version.

We introduce the \emph{dual variables}:
\begin{equation}
  \omega_J(s),\quad \nu_J(s), \quad \alpha_{n}
\end{equation}
where $\omega_J(s)$ has support wherever the Roy equations are imposed, $\nu_J(s)$ wherever the unitarity inequality is imposed, and $\alpha_{n}$ is non-zero only for $n$ odd. A simple exercise in semidefinite progamming duality shows that, if we impose:\footnote{We henceforth leave implicit the dependence of the kernels $R_{\mu \, \ell}^{(J)}$ and $R_{\mu\, \ell}^{[n]}$ on the subtraction point $(s_0, t_0)$.} 
\begin{align}
  \int ds \, \omega_0(s) &= 0 \label{dualequality}\\
  v_p + \sumint\limits_{J,s}\, \omega_J(s) R^{(J)}_{\mu_p \, \ell_p} (s) + \sum_{n \text{odd}} \alpha_{n} R^{[n]}_{\mu_p,\ell_p}(t_c) &\leq 0 \label{dualinequality}\\
  \begin{pmatrix}
  \nu_J(s) & \omega_J(s)/2\\
  \omega_J(s)/2 & \nu_J(s) - \xi_J(s)
  \end{pmatrix}&\succeq 0\,\label{dualpositivedefiniteness},
\end{align}
with
\begin{equation}
  \xi_J(s) \colonequals \sumint\limits_{\ell,v}\, \omega_\ell(v) R^{(\ell)}_{s \, J} (v)+ \sum_{m,n} \alpha_{m,n} R^{[n]}_{s\, J}(t_c)\,,
\label{xi_J}
\end{equation}
then simply combining all the constraints produces: 
\begin{equation}
\label{dualitygap}
  \sum_p v_p g_p^2 \leq \sumint\limits_{J,s}\, \frac{2 \nu_J(s)}{\tilde \rho_s}\,.
\end{equation}
Therefore, any set of dual variables that obeys the constraints imposes an upper bound on the primal objective. This bound remains rigorously valid even if we truncate the space of dual variables as we will do below.

It is however important that the `dual positivity conditions' \eqref{dualpositivedefiniteness} must be obeyed for all physical $s$ and $J$. To see this we recall that free primal variables become Lagrange multipliers for dual constraints. Therefore, forgetting the \emph{dual} positivity constraints in some region is tantamount to setting the \emph{primal} partial waves $f_J(s)$ to zero in the same region. This would be unphysical: the results on the primal side would be artificially weak, whereas the dual bounds would be too strong.

Of course, outside the support of $\omega_J(s)$ and $\nu_J(s)$ the dual positivity equations reduce to the simple linear inequality $\xi_J(s) \leq 0$. Similarly they reduce to $\nu_J(s) \geq \max(0,  \xi_J(s))$ whenever $\nu_J(s)$ has support but $\omega_J(s)$ does not. And lastly they imply that it is not meaningful to give $\omega_J(s)$ support wherever $\nu_J(s)$ is set to zero.

\section{Implementation details}
We minimize the right-hand side of equation \eqref{dualitygap} subject to the constraints given in equations \eqref{dualequality}, \eqref{dualinequality}, and \eqref{dualpositivedefiniteness}. We use SDPB \cite{Simmons-Duffin:2015qma,Landry:2019qug} to solve the semidefinite program numerically with the following choices for the various functions and parameters:
\begin{itemize}
  \item The imposition of the Roy equations. We chose $\omega_\ell(s)$ to be non-zero only for $\ell = 0, 2, 4, \ldots L_\text{max}$ and for $4 \leq s \leq \mu^2$. We pick $\mu^2 = 12$ since any larger value runs into difficulties with the dual positivity constraints at large $J$. (This rather unintuitive result is derived in appendix \ref{app:asymptotics}.) Within this domain we use an essentially polynomial basis as explained in appendix \ref{app:functionsansatze}. We truncate the basis to include $P+1$ terms.
  \item The imposition of the primal unitarity equations, as captured in an ansatz for $\nu_\ell(s)$. We chose $\nu_\ell(s)$ to again be non-zero only for $\ell = 0, 2, 4, \ldots L_\text{max}$, but now include all $4  \leq s < \infty$. An essentially polynomial ansatz for $\nu_\ell(s)$ for $4 \leq s \leq \mu^2$ contains $N+1$ terms and for  $\mu^2 \leq s < \infty$ we include $M+1$ terms, see again appendix \ref{app:functionsansatze} for details.
  \item The choice of $t_c$ and the number of crossing equations to impose around it, corresponding to the number of non-zero $\alpha_n$. We kept only one term, corresponding to $\alpha_{1}$, at $t_c = \frac{2}{3}$ . (In future studies one could also include derivatives to $R^{[n]}_{s\,J}(t_c)$ with respect to $t_c$, or sample at multiple values of $t_c$.)
  \item The subtraction point $(s_0, t_0) = (2,0)$. 
\end{itemize}
We emphasize that the bounds we obtain will be fully rigorous even when $\omega_\ell(s)$, $\nu_\ell(s)$ and $\alpha_{n}$ are truncated as above. In contrast, limits on computational resources also forces us to choose:
\begin{itemize}
  \item The discretization of the dual positivity equations. The space of physical $s$ and $J$ splits into three regions: for low $J$ and $s$ both $\omega_J(s)$ and $\nu_J(s)$ have support, for low $J$ and large $s$ only $\nu_J(s)$ has support, and for high $J$ and any $s$ both are set to zero. We discretized $s$ in the first two regions with 200 points and in the third with 400 points and further truncated spins up to $J_\text{max} = 32$. We found experimentally that including a finer grid in $s$ or larger $J$ would not meaningfully change our bounds. In appendix \ref{app:asymptotics} we analyze dual positivity for large $J$ and for large $s$ behavior. Our main conclusion is that dual positivity \emph{simplifies} in these limits. Indeed with our choice of dual variables it can be satisfied for all $s$ and $J$.
\end{itemize}

Below we will plot our results as a function of $L_\text{max}$. For each fixed $L_\text{max}$ we increased $M$, $N$ and $P$ until the bounds no longer depended on them, which turned out to be the case for $M = 40$, $N=10$, and $P=20$.

\section{Amplitude Reconstruction}
One can construct unitarity-saturating partial waves from the dual variables:
\begin{equation}
  1 + i \tilde \rho_s f_J(s) = - \frac{\xi_J(s) + i \omega_J(s)}{\sqrt{\xi^2_J(s) + \omega^2_J(s) }}
\end{equation}
Although this works quite generally, it is also the correct `extremal' partial wave when both the primal and dual positive semidefiniteness constraints are saturated. In the domain where we do not impose the Roy equations we set $\omega_J(s)  = 0$ and then $1 + i \tilde \rho_s f_J(s) = \pm 1$. And if we neither impose the Roy equations nor unitarity then $\xi_J(s) \leq 0$ necessarily, so only the plus sign survives.

Given the implementation details of the previous section, we will obtain non-trivial unitarity-saturating partial waves only for $4 \leq s \leq \mu^2 = 12$ and $\ell = 0, 2, 4,\ldots L_\text{max}$. For $s < \mu^2$ and still $\ell < L_\text{max}$ the phase shifts can either be 0 or $\pi$, and for $\ell > L_\text{max}$ the phase shifts are all set to 0. This structure of the extremal solution does not take away from the fact that our bounds are rigorous and apply to all scattering amplitudes. Fundamentally this is because we truncated the dual problem instead of the primal problem.

\begin{figure}[t]
    \centering
    \includegraphics[width=\linewidth]{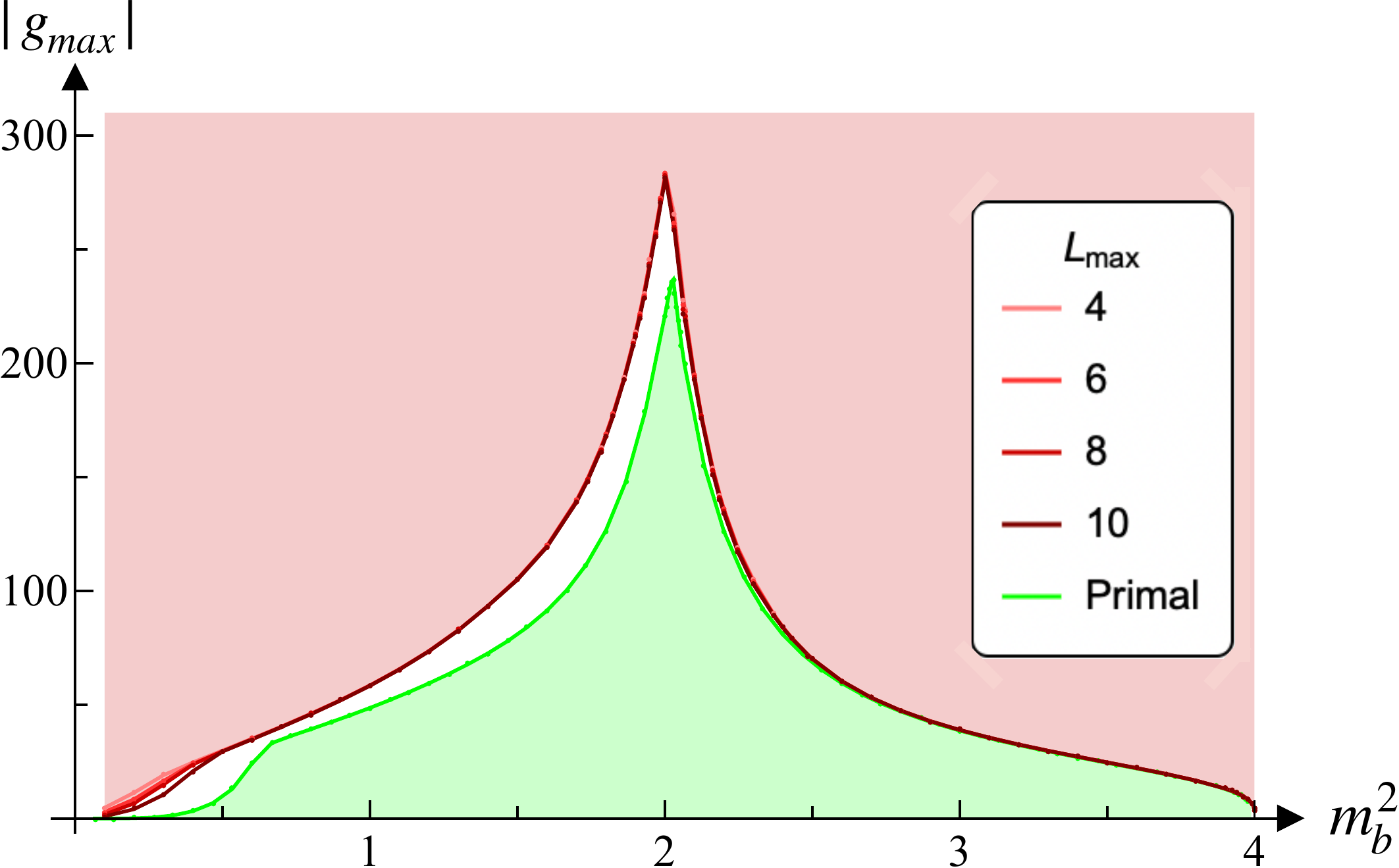}
    \caption{Bounds on the maximum residue at a scalar bound state pole of mass $m_b^2$. In green, the primal bound obtained by constructing maximal analytic, crossing, and unitary amplitudes. In red, the rigorous dual excluded region.}
    \label{fig:scalar_pole}
\end{figure}

\section{Dual Rigorous Bounds}
\subsection{Single scalar bound state}
First we study the maximum coupling of a single scalar bound state of mass $m_b^2$. In \cite{Paper3} the primal version of this problem was introduced, which amounted to scanning the space of crossing symmetric amplitudes obeying maximal analyticity. The maximum coupling \footnote{The absolute value sign is necessary because the setup is insensititve to the sign of the coupling, and more generally only relative signs like that of $g_{112}/ g_{222}$ are physically meaningful.} $|g_\text{max}|$ as a function of $m_b^2$ obtained in this way is plotted in figure \ref{fig:scalar_pole} in green. In the same figure, in shades of red, we plot our new dual upper bound obtained by minimizing the right-hand side in~\eqref{dualitygap}. For $m_b^2 > 2$ we find quantitative agreement, but for lighter bound states there is a finite region between the two boundaries. This gap does not seem to disappear by improving the numerics, since both primal and dual appear to have converged reasonably well.

This gap is likely due to the different constraints imposed for the two different methods. The dual bounds obtained using our algorithm are rigorous but conservative, since they use only proven analyticity \cite{Martin:1965jj}. Moreover, the primal ansatze of \cite{Paper3} cannot describe amplitudes with growing cross sections at high energies.

In Appendix~\ref{scalar_pheno} compare the extremal amplitudes saturating the primal bounds against the phase shifts reconstructed from the dual setup. We in particular show that the size of the gap is correlated to higher spin dominance in the region not covered by the Roy equations.

\begin{figure}[t]
    \centering
    \includegraphics[width=\linewidth]{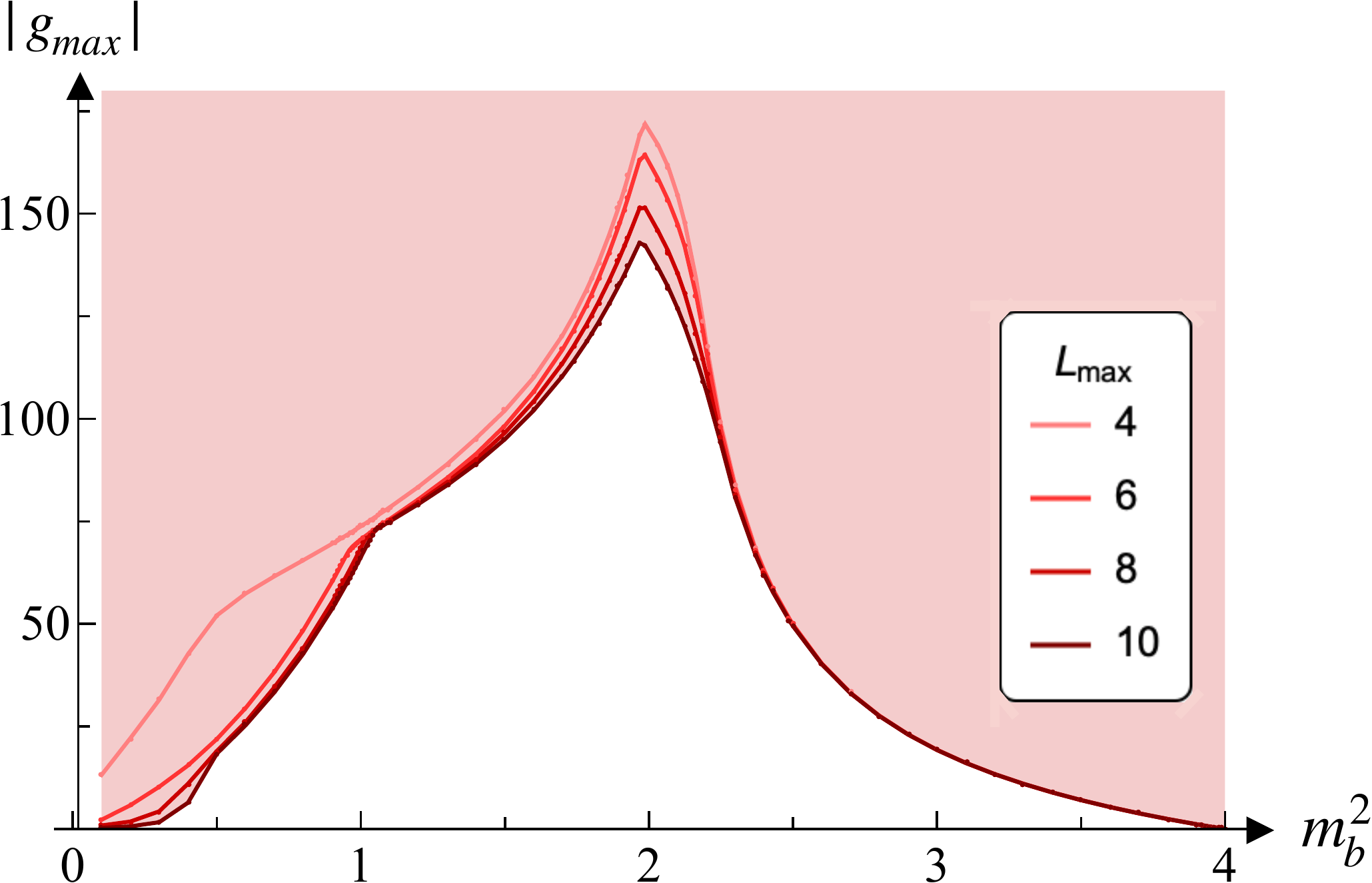}
    \caption{Bound on the maximum residue $|g_\text{max}|$ at the spin-two bound state of mass $m_b^2$. The red region is rigorously excluded. Different colours correspond to a different number of constraints.}
    \label{fig:spin2_pole}
\end{figure}

\begin{figure}[t]
    \centering
    \includegraphics[width=\linewidth]{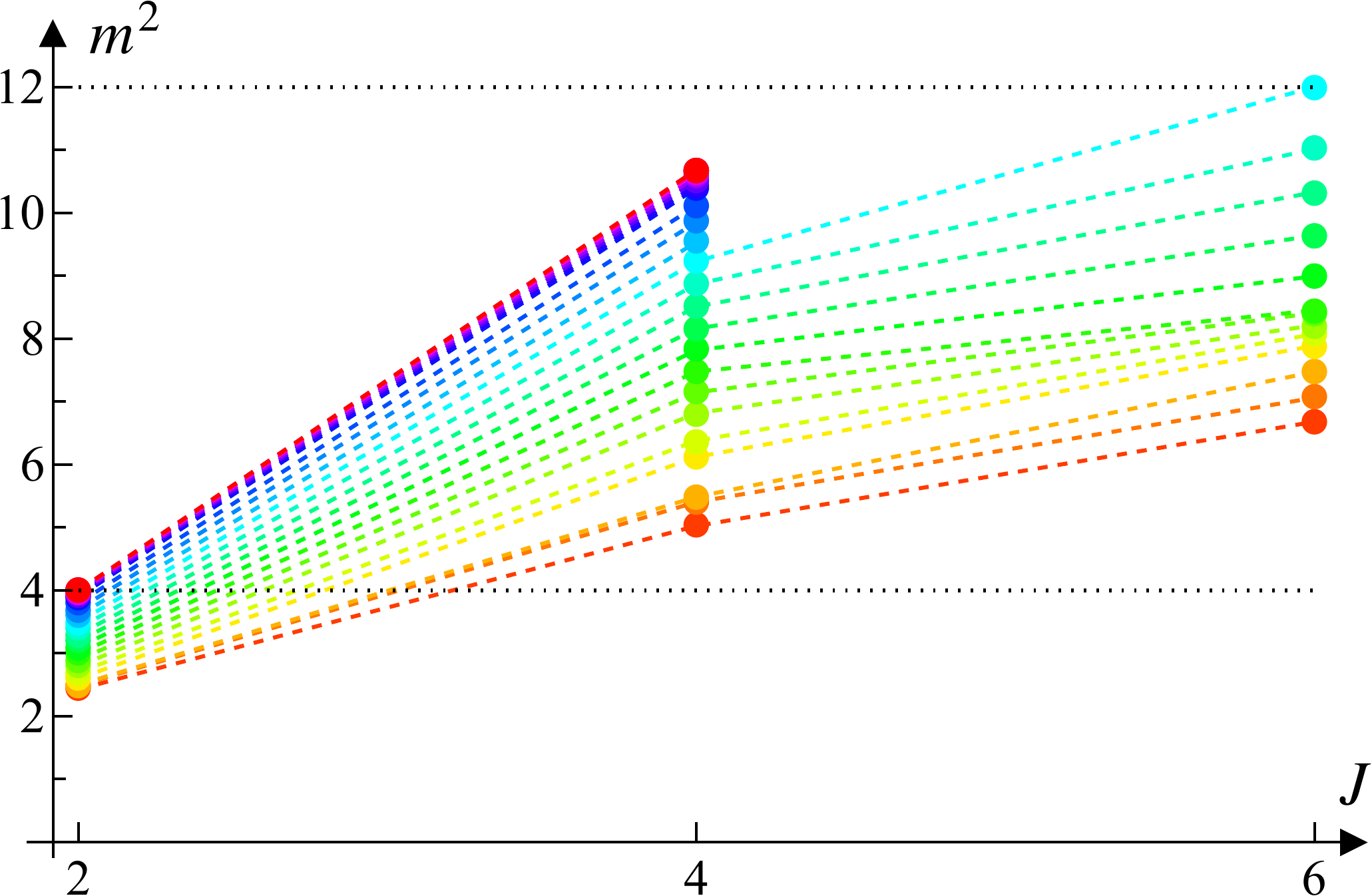}
    \caption{Mass of the resonance $m^2$ as a function of the spin $J$ extracted from the extremal amplitudes in figure~\ref{fig:spin2_pole} for $m_b^2>2$. Resonances extracted from the same amplitude are denoted with the same colour. Dashed lines delimitate the window where we impose Roy equations and extract the resonances.}
    \label{fig:Regge_plot_spin2pole}
\end{figure}

\begin{figure}[t]
    \centering
    \includegraphics[width=0.9\linewidth]{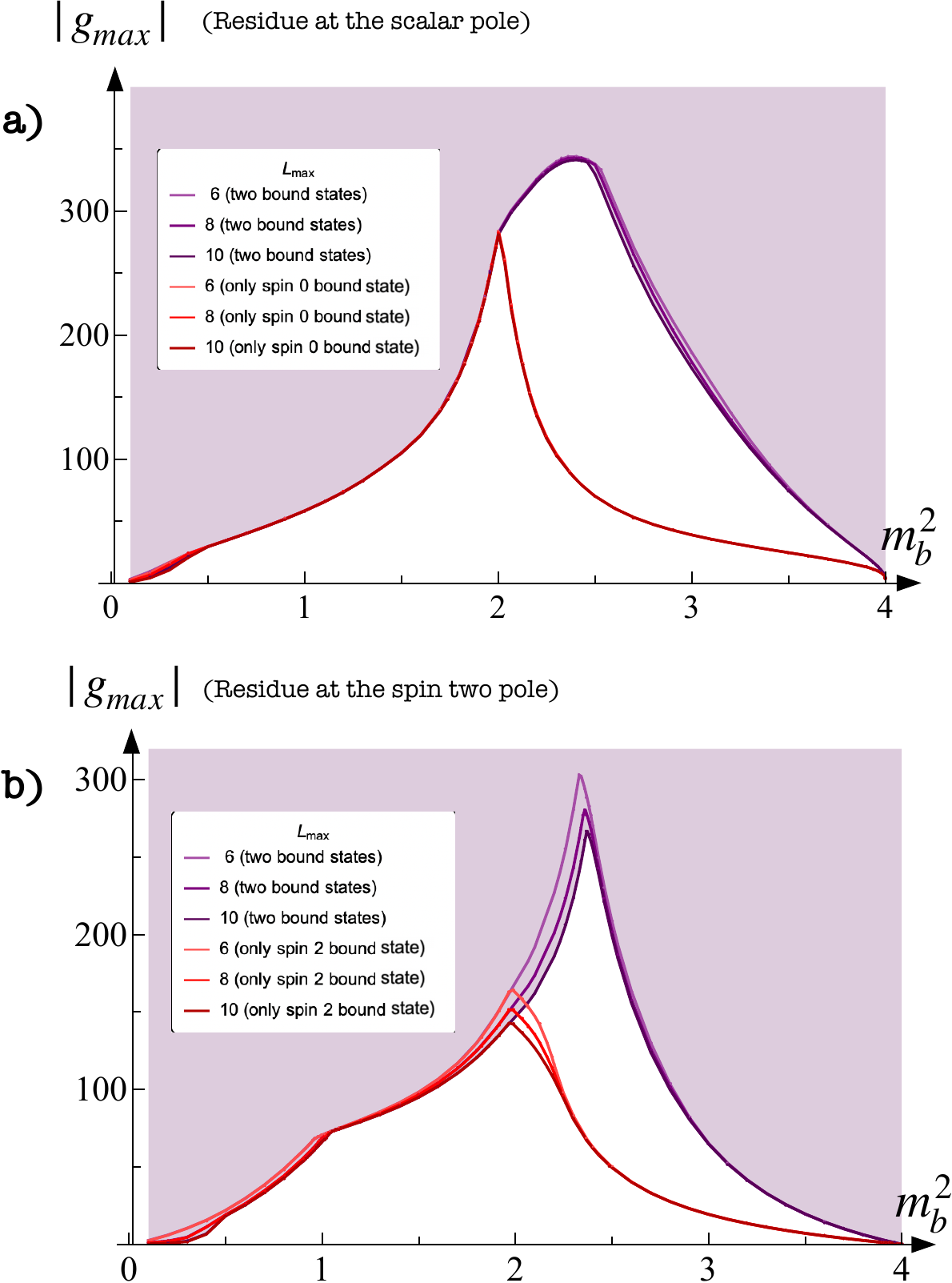}
    \caption{In purple, bound on the residue at the scalar bound state, panel $a)$, and at the spin-two bound state, panel $b)$, in presence of the self-coupling pole with mass $m=1$. In red, the same bound in absence of the self-coupling pole given respectively in fig.~\ref{fig:scalar_pole}, and in fig.~\ref{fig:spin2_pole}.}
    \label{fig:two_bound_states}
\end{figure}

\begin{figure*}[t]
    \centering
    \includegraphics[width=\linewidth]{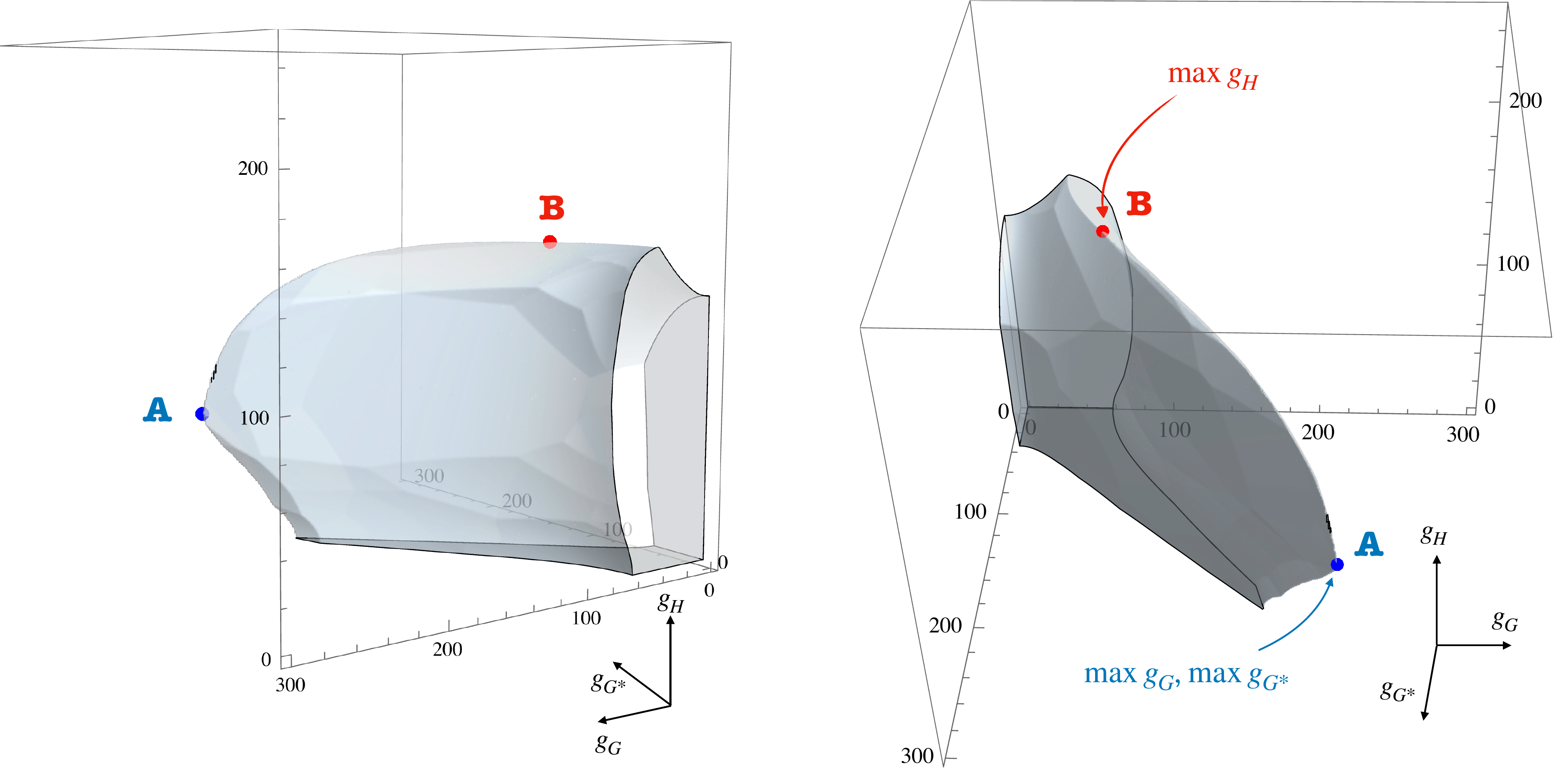}
    \caption{The \emph{glue-hedron}. Glueball couplings for SU(3) YM must be contained inside this 3d space. The black lines denote the boundaries of the glue-hedron on the planes where one of the couplings is zero.}
    \label{fig:glue-hedron}
\end{figure*}

\subsection{Single spin-two bound state}
Next we consider the case of a single spin-two bound state pole in the amplitude. In figure~\ref{fig:spin2_pole}, in red, we show the excluded region as a function of the number of Roy equations $L_\text{max}$. We see a first kink at $m_b^2$ slightly bigger than 1, for which we unfortunately do not have a good explanation. The cusp at $m_b^2$ slightly smaller than 2 is reminiscent of a divergence at $m_b^2 = 2$ in the analogous bound in two dimensions \cite{Paper2}. In that case the $s$- and $u$-channel poles overlap perfectly and cancel each other out, and presumably a similar but imperfect \emph{screening} \cite{Homrich:2019cbt} behavior occurs here.

Let us comment on the higher-spin resonances that we observed in the extremal amplitudes, deferring a more general discussion of their physics to appendix \ref{sec:phenospin2}. We estimated the masses of these resonances for spin $J = 4, 6$ by plotting the location where the phase shift $\delta_J(m^2)=\pi/2$ and depicted the results in figure~\ref{fig:Regge_plot_spin2pole}. Each color corresponds to a particular choice of $m_b^2$, which in the figure is just the value at $J=2$.

We observe that particles in the same amplitude organize into nearly linear Regge trajectories. (It is unclear to us whether the non-linearity is physical or rather a numerical artefact.) We also see that the spin four particle approaches the threshold as the spin two bound state mass approaches $m_b^2=2$ from above. This gives a partial explanation for the kink observed there: if we would continue the trend then the spin-four resonance would become stable and show up as a singularity in the physical sheet, but this is forbidden by our working hypothesis.

\subsection{Two poles}
So far we considered the exchange of only a single non-trivial bound state, corresponding to a pole in the amplitude only at $s, t, u = m_b^2$. In this subsection we will include a self-coupling of the external particle, \emph{i.e.} we include another pole in the amplitude at $s, t, u = 1$. We demand that the residue of this pole is non-negative and then again bound the maximal coupling to the new bound state at $m_b^2$.

The results of this study are shown in purple in figure \ref{fig:two_bound_states}a for a scalar bound state and in figure \ref{fig:two_bound_states}b for a spin two bound state. Compared to the results of the previous subsections (in red), the bound on the maximal coupling remains unchanged for $m_b^2 \lesssim 2$ and becomes weaker for $m_b^2 \gtrsim 2$.

We can qualitatively understand this result by appealing to the screening phenomenon found and discussed earlier in \cite{Paper2,Homrich:2019cbt,Guerrieri:2020kcs}. Consider the amplitude $T(s,t)$ in the forward limit $t=0$. Crossing symmetry becomes $T(s,0) = T(4-s,0)$ so it suffices to consider the half-plane with $\text{Re}(s) > 2$. Now, for $m_b^2 > 2$ the two poles in that half-plane at $s = m_b^2$ and $s = 3$ and have residues with opposite signs, allowing for partial cancellation of their effect on the physical region which sits at $s \geq 4$. This is why adding an extre pole is expected to lead to a strictly weaker bound in this region. On the other hand, for $m_b^2 < 2$ their residues are of the same sign, so extremizing one coupling would naturally set the other to zero. The bound then reduces to a single-particle bound, which is what we observe in figure \ref{fig:two_bound_states}.

\subsection{The glue-hedron}
We are now ready to consider the $GG \to GG$ amplitude. Besides allowing for a self-coupling pole at $m^2 = 1$ in the amplitude, we also include the poles corresponding to the three other stable particles listed in table~\ref{SpectrumSU3}. For simplicity we fix their masses to the central values and do not take into account the uncertainty in the lattice determination. Our analysis is however easily repeated for other masses if this turns out to be necessary.

In the first row of table \ref{max_couplings} we report the absolute upper bound for each of the couplings found at the different poles. The second row of the table shows the same bounds with the excited spin-two glueball $H^*$ removed from the set of bound states. This glueball is very close to threshold and could in actuality be unstable, but if this is so then the bounds for the other couplings strengthen, albeit only by a few percent.

\begin{table}[h]
  \centering
  \begin{tabular}{| c | c | c | c|} 
   \hline
  \quad $\max |g_G|$ \quad & \quad $\max |g_H|$ \quad & \quad $\max |g_{G^*}|$ \quad & \quad $\max |g_{H^*}|$ \quad \\ 
   \hline
   \hline
  213 & 158 & 224 &  2.15 \\
   \hline
  206 & 156 & 217 &  -- \\
   \hline
  \end{tabular}
  \caption{Upper bounds on glueball three-point couplings, either with (first row) or without (second row) the $H^*$.}
  \label{max_couplings}
\end{table}

To get an idea of the strength of these bounds, we note that \cite{DEFORCRAND1985107} used lattice results to estimate $g_G \approx 50 \pm 7$ for $SU(3)$ Yang-Mills theory.\footnote{The quantity $G=155\pm 45$ measured in \cite{DEFORCRAND1985107} is related to our coupling $g$ by $G=3g^2/(16\pi)$.} This interval would partially be excluded if the $G$ would be the only bound state, see figure \ref{fig:scalar_pole} at $m_b^2 = 1$. However the actual bound on $|g_G|$ in table \ref{max_couplings} is much weaker and easily allows this first lattice estimate. We can also intuitively explain the relative weakness of this bound: as shown in figure \ref{fig:poles} the $u$-channel pole of $G^*$ lies almost at $s = 1$ and so the two poles approximately cancel each other. Below we will indeed see that $g_{G^*}$ is maximized whenever $g_G$ is.

In figure~\ref{fig:glue-hedron}, we show the \emph{glue-hedron}: the allowed region in the three-dimensional space spanned by $\{ g_G, g_G^*,g_H\}$. (It is convex in the space $\{ g_G^2, g_{G^*}^2,g_H^2\}$ because the space of allowed scattering amplitudes is.) To obtain it we left $g_{H^*}$ free and extremized the linear combinations $\vec n \cdot \{g_G, g_{G^{*}}, g_H\}$ for 263  different three-dimensional unit-norm vectors $\vec n$. If $SU(3)$ Yang-Mills theory obeys (essentially) the Wightman axioms then its glueball couplings must lie inside the glue-hedron.

We can identify the extremal couplings in table \ref{max_couplings} with two points of the glue-hedron as indicated in figure \ref{fig:glue-hedron}. There is one point A that maximizes both $g_G$ and $g_{G^*}$ for some finite but non-extremal $g_H \approx 65 $. (Closer inspection shows that $g_{H*}$ is in fact also maximized at this point.) The maximum of $g_H$ is attained at another point, which we call B. Here $g_{G^*}\approx 71$ and $g_G\approx 61$ are both non-extremal. We will now proceed to analyze the phase shifts at these points.

\section{Extremal amplitudes}
\begin{figure}[t]
    \centering
    \includegraphics[width=0.8\linewidth]{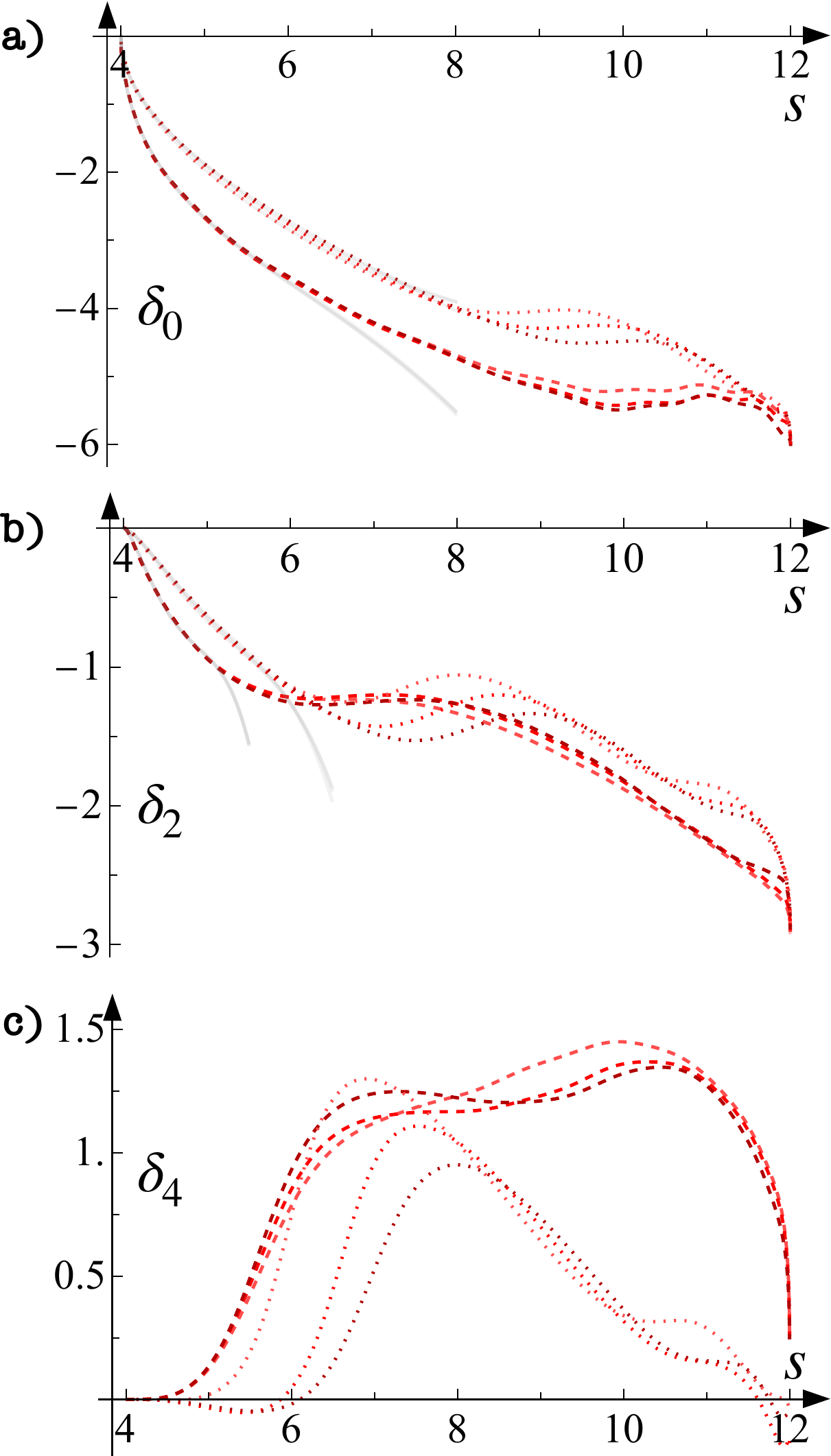}
    \caption{Phase shifts for the extremal cusps A (dashed) and B (dotted). The different shadings correspond to different $L_\text{max}$, from 6 (lightest) to 8 and then to 10 (darkest). The grey curves are polynomial fits to estimate the low-energy data.}
    \label{fig:phase_shifts_cusps}
\end{figure}
In figure~\ref{fig:phase_shifts_cusps} we plot the phase shifts $\delta_\ell$ at the A (dashed) and the B cusp (dotted) for spin $0$, $2$, and $4$. We recall that the extremal partial waves that we obtain saturate unitarity by construction, so the phase shifts are real, and that they are non-trivial only for $0 \leq s \leq \mu^2 = 12$.

The spin 0 phase shifts $\delta_0$ can be parametrized at threshold as
\begin{equation}
k_s\cot\delta_0=\frac{1}{a_0}+\frac{1}{2}r_0^2 k_s^2+\dots,
\label{eq:definition_scatt_length}
\end{equation}
where $k_s=\sqrt{s-4}/2$, $a_0$ is the scattering length and $r_0$ the effective range. Using a standard threshold expansion \cite{Correia:2020xtr} which involves three free parameters, we can estimate:
\begin{eqnarray}
A:\qquad a_0\approx -5.9, \quad r_0\approx 3.8\\
B:\qquad a_0\approx -3.7, \quad r_0\approx 2.6
\end{eqnarray}
Our fits are shown as the gray curves in figure \ref{fig:phase_shifts_cusps}. Note that the scattering length are significantly below the minimum value $\min a_0=-1.75$ that can be reached in the absence of poles~\cite{Lopez:1975ca}. 

For the spin $2$ phase shift we obtain a good fit with the expansion:
\begin{equation}
  k_s^3 \cot\delta_2=\frac{1}{\widetilde a_2}+\frac{1}{2}\widetilde r_2^2 k_s^2+\dots;
\end{equation}
to obtain:
\begin{eqnarray}
  A:\qquad \widetilde a_2\approx -23, \quad \widetilde r_2\approx -0.4\\
  B:\qquad \widetilde a_2\approx -10, \quad \widetilde r_2\approx -0.5
\end{eqnarray}
Now, however, the coefficients $\widetilde a_2$ and $\widetilde r_2$ are not the usual physical scattering length and effective range. Indeed, the above expansion corresponds to a partial wave of threshold behaviour the form
\begin{equation}
  f_2(k_s)=k_s^2\left(f_2^{(2)} +f_2^{(4)} k_s^2+\frac{i}{2} (f_2^{(2)})^2 k_s^3+\dots\right).
  \label{fit_test}
\end{equation}
Such behaviour is compatible with unitarity, nevertheless it would be interesting to find a quantum mechanical potential to model this behaviour.~\footnote{Spin-two partial amplitudes usually behave as $f_2(k_s)\sim k_s^4$, as $k_s\to 0$. }

In figure~\ref{fig:phase_shifts_cusps}c we look at the higher spin wave $\ell=4$. It is entertaining to compare this with $SU(3)$ Yang-Mills theory, where we would expect a spin four resonance with mass $m_4^2\approx5$~\cite{Athenodorou:2020ani,Athenodorou:2021qvs}. We indeed observe a pretty large phase shift there for cusp A, but not a clear sign of this resonance. Of course there is no a priori reason that $SU(3)$ Yang-Mills theory must live at the cusp.

\section{Discussion and Outlook}
In this letter, we generalized the dual S-matrix bootstrap method to constrain three point couplings to bound state particles of any spin. We then demonstrated the method for simple cases of one and two bounds states of spin 0 and spin 2, along the way verifying the estimates of  the maximal coupling to spin 0 bound state presented in \cite{Paper3}. However, the main result of this letter is the glue-hedron in figure \ref{fig:glue-hedron}, which we obtain by bounding the space of three-point couplings in pure SU(3) Yang-Mills.

Two future sources of improvement immediately stand out. Firstly, we could consider the mixed system of scattering of the lightest glueball along with one or more of the heavier glueballs. Since the mass ratios here are greater than $\sqrt{2}$ in units of the lightest glueball, this would entail dealing with anomalous thresholds -- see~\cite{Correia:2022dcu} for a recent discussion. In addition, the particle $H$ has spin 2 and therefore we would also have the complication of spinning external particles~\cite{Hebbar:2020ukp}. Secondly, we could use additional input from lattice measurements, for example by also including the scattering length as an input along with the glueball masses. As these more complicated studies shrink the allowed space of glueball couplings further and further, one hopes that the extremal phase shifts from the bootstrap will start resembling actual SU(3) Yang-Mills scattering phase shifts.

Our generalization of the dual S-matrix bootstrap to include bound states allows us to constrain amplitudes in many different ways for a wide variety of physical systems. Besides the bound-state couplings, one can constrain directly observable data such as scattering lengths, effective ranges, or more generally the low-energy behavior of the partial waves. On the longer term it would be great if the numerical results would inspire a better \emph{analytic} understanding of the bounds and the structures we observe in the extremal amplitudes, but this may require substantial advances in the complex analysis of functions of two variables.

\section*{Acknowledgements}
We thank Mattia Bruno, Miguel Correia, Maxwell Hansen, Joao Penedones and Pedro Vieira for useful discussions and suggestions. AH and BvR are supported by Simons Foundation grant \#488659 for the Simons Collaboration on the non-perturbative bootstrap and by the European Union (ERC, QFTinAdS, 101087025). Research
at the Perimeter Institute is supported in part
by the Government of Canada through NSERC and by
the Province of Ontario through MRI. AG is supported
by the European Union - NextGenerationEU, under the
programme Seal of Excellence@UNIPD, project acronym
CluEs. Views and opinions expressed are those of the author(s) only and do not necessarily reflect those of the European Union or the European Research Council Executive Agency. Neither the European Union nor the granting authority can be held responsible for them.

\appendix
\section{Dispersion relation derivation}
\label{app:dispersionderivation}
We consider the amplitude $T(s,t)$ for the scattering of scalars of mass $m=1$. For a fixed $t, t_0$ we can write:
\begin{multline}
  \frac{1}{2} \left( T(s,t) - T(t_0,t) + T(4-s-t,t) - T(4-t_0-t,t) \right)\\
  = \frac{1}{2} \oint_C \frac{dv}{2\pi i} T(v,t) K(v,s,t;t_0)
\end{multline}
with the kernel
\begin{equation}
  \la{kernel}
K(v,s,t;t_0)= \frac{1}{v{-}s}+\frac{1}{v{-}4{+}s{+}t}-\frac{1}{v{-}t_0}-\frac{1}{v{-}4{+}t{+}t_0}\,,
\end{equation}
and with $C$ a contour in the $v$ plane that encircles the four points $\{s, t_0,4-s-t,4-t_0-t\}$ but avoids any singularities in $T(v,t)$. Next we blow up this contour, and since $K(v,s,t;t_0) = O(|v|^{-3})$ as $|v| \to \infty$ we can drop the arcs at infinity because of equation \eqref{Tasymptoticbehavior}. Using also crossing symmetry $T(s,t) = T(4-s-t,t)$ we get:
\begin{equation}
  T(s,t) - T(t_0,t) = \int \frac{dv}{\pi} T_v(v,t) K(v,s,t;t_0)
\end{equation}
with $T_v(v,t)= \frac{1}{2i} \left(T(v+ i\epsilon,t) - T(v-i\epsilon,t)\right)$. The integration range covers all values of $v$ where the $s$-channel discontinuity is non-zero, which in our case includes some isolated poles and then the cut with $v \in [4,\infty)$.

We can add to this the equivalent relation written for $T(t,t_0) - T(s_0,t_0)$, and using $T(t,t_0) = T(t_0,t)$ we obtain:
\begin{multline}
T(s,t) = T(s_0,t_0) + \sum_{p \in \cP} g^2_p R_{\mu_p^2\, \ell_p}(s,t;s_0,t_0) \\
+\frac{1}{\pi}\int\limits_{4}^\infty dv\, \[T_v(v,t) K(v,s,t;t_0){+} T_v (v,t_0) K(v,t,t_0;s_0)\]\,,
\label{disp}
\end{multline}
Here
\begin{multline}
R_{\mu^2\, \ell}(s,t;s_0,t_0) = P_\ell\left(1 + \frac{2t}{\mu^2 -4}\right) K(\mu^2,s,t;t_0) \\
+ P_\ell\left(1 + \frac{2t_0}{\mu^2 -4}\right) K(\mu^2,t,t_0;s_0)
\end{multline}
is the contribution of a pole in the amplitude of the form:
\begin{equation}
  T(s,t) \supset - \frac{g_p^2 P_\ell\left(1 + \frac{2t}{\mu^2 -4}\right)}{s - \mu^2}
\end{equation}
which corresponds a bound-state particle of mass squared $\mu$ and spin $\ell$. We introduce the partial wave decompositon
\begin{equation}
  \label{partialwavedecompositionT}
  T(s,t) = \sum_{\ell} n_\ell^{(d)} f_\ell(s)  P_\ell^{(d)}\left(1 + \frac{2t}{s -4}\right)
\end{equation}
with $\ell$ even and $n_\ell^{(4)} = 16 \pi (2\ell+1)$. This allows us to write equation \eqref{bestdispersion} in the main text, with the shorthand:
\begin{equation}
\label{defsumint}
  \sumint\limits_{\ell,v} \, \Xi_{\ell}(v) = \frac{1}{\pi} \int\limits_{4}^\infty dv \sum_\ell n_\ell^{(d)} \Xi_{\ell}(v) \,.
\end{equation}

\section{Functional ansatze}
\label{app:functionsansatze}
In this appendix we discuss the space of functions from which we sample $\omega_L(s)$ and $\nu_L(s)$. We will take them to be non-zero only for even $L \leq L_\text{max}$. The support of $\omega_L(s)$ is further restricted to $s \in [4, \mu^2]$ whereas $\nu_L(s)$ is allowed to be nonzero for all $s \in [4,\infty)$.

We use the maps%\BvR{What is $\sigma$?}
\begin{equation}
\begin{split}
  s^{\rm IR}(z)&=\frac{\mu^2-4}{2}z+\frac{\mu^2+4}{2}\\
  s^{\rm UV}(z)&=2\mu^2-\sigma+\frac{2(\mu^2-\sigma)}{\sin(\pi z/2)-1}  
\end{split}
\end{equation}
to parametrize respectively the interval $[4,\mu^2]$, and $[\mu^2,\infty)$ in terms of a variable $z\in [-1,1]$, where we chose $\sigma = 20$. We then use the ansatz
\begin{equation}
 \left(\frac{d s^{\rm IR}(z)}{dz}\right) \omega_L(z)= \sum_{p=0}^P c^{(\mu)}_{L\,p} T_p(z)
\end{equation}
with $T_p(\cos(\theta)) \colonequals \cos(p \theta)$ the Chebyshev polynomials of the first kind. The derivative on the left-hand side facilitates integration over $s$, although here it is merely a constant term. For $\nu_L(s)$ at high energies we will use
\begin{multline}
  s \in [\mu^2,\infty]: \quad  \left(\frac{d s^{\rm UV}(z)}{dz}\right) \nu_L(z) = \sum_{m=0}^M c^{(\nu) \rm UV}_{L\,m} T_m(z)\,,
\end{multline}
and for low energies our ansatz reads
\begin{multline}
  s \in [4,\mu^2]: \quad  \left(\frac{d s^{\rm IR}(z)}{dz}\right) \nu_L(z) =\\
  \frac{\tilde c_{0}^{(\nu)} \delta_{L,0}}{\sqrt{x+1}}+\sum_{n=0}^N c^{(\nu) \rm IR}_{L\,n} T_n(z)\,,
\end{multline}
with an extra term $\tilde c_{0}$ which allows for the extremal partial wave to have a threshold scalar bound state.

Altogether we have a discrete set of degrees of freedom spanned by the coefficients
\begin{equation}
  c_{L\, p}^{(\mu)},\quad c_{L\, m}^{(\nu) \rm UV},\quad  c_{L\, m}^{(\nu) \rm IR},\quad \tilde c^{(\nu)}_0\,,
\end{equation}
which makes the problem suitable for a numerical approach. Our results can only improve if we increase the parameters $P$, $M$, $N$ and the maximal spin $L_\text{max}$ for which we take $\omega_L(s)$ and $\nu_L(s)$ to be non-zero.

\section{Asymptotics}
\label{app:asymptotics}
In this appendix we verify that the dual positivity conditions can be satisfied for very large $J$ and $s$ even with a finite-dimensional ansatz.

In the following we will make essential use of our choice of $t_0 = 0$ in the subtraction point of the dispersion relation, and likewise we will suppose that we check crossing symmetry by taking derivatives around a point such that $t_c \in (0,2)$.

\subsection{Large spin}
For very large $J$ we are outside the support of $\omega_J(s)$ and $\nu_J(s)$ so the dual positivity constraint \eqref{dualpositivedefiniteness} reduces to
\begin{equation}
  0 \geq \xi_J(s) = \sumint\limits_{\ell,v}\, \omega_\ell(v) R^{(\ell)}_{s \, J} (v)+ \sum_{n = 1,3,5, \ldots n_\text{max}} \alpha_{n} R^{[n]}_{s\, J}(t_c)\,,
\end{equation}
where we recalled the definition of $\xi_J(s)$ given in the main text. The large $J$ asymptotics of the two $R^{(\cdot)}_{s\, J}$ are easily determined from the asymptotic behavior of the Legendre polynomials. In $d=4$ we have, for $\theta > 0$,
\begin{equation}
  P_J(\cosh(\theta)) = \frac{e^{J \theta}}{ \sqrt{\pi J (1 - e^{-2\theta})}}  \left( 1 + O(J^{-1})\right)
\end{equation}
Below we will use this to write
\begin{equation}
  \begin{split}
    \frac{d}{d \theta} P_J(\cosh(\theta)) &= J P_J(\cosh(\theta)) \left( 1 + O(1/J) \right)\\
    \int^{\theta} d\theta' \, P_J(\cosh(\theta')) &= \frac{1}{J} P_J(\cosh(\theta)) \left( 1 + O(1/J) \right)
  \end{split}
\end{equation}
which is valid in any dimension. Since we chose the subtraction point such that $t_0 = 0$, we find:
\begin{multline}
    R^{[n]}_{s\, J}(t_c) =  K(s, 4- 2 t_c, t_c;0) P_J\left( 1 + \frac{2 t_c}{s-4}\right) \times\\
    J^n \left(\frac{8 t_c}{(s-4)^3} \left( 1 + \frac{2 t_c}{s-4}\right)\right)^{n/2}\left( 1 + O(1/J)\right)
\end{multline}
For $0 < t_c < 2$ the kernel $K(s, 4 - 2 t_c, t_c;0)$ is smooth and negative for all $s \geq 4$. All the other factors in the above expression are manifestly positive, which is good news: provided no further exponentially growing terms occur (see however below), the simple constraint
\begin{equation}
  \label{largeJalphaconstraint}
  \alpha_{n_\text{max}} > 0
\end{equation}
would suffice to ensure dual positivity for all $s \geq 4$ in the asymptotically large $J$ regime.

For $R^{(\ell)}_{s\, J}(v)$ we find that the large $J$ behavior is dominated 
by the behavior of the integral as $z$ approaches $0$ and $v$ approaches its maximal value $\mu^2$. It takes the form:
\begin{multline}
  \label{largeJRell}
  \sumint\limits_{\ell,v}\, \omega_\ell(v)R^{(\ell)}_{s \, J} (v) = K(s,\mu^2, -(\mu^2 - 4)/2;0) \times \\
  \frac{\mu^2 + 4 - 2 s}{\pi J^2} P_J\left(1 - \frac{\mu^2 - 4}{s-4}\right) \times \\
  \sum_\ell (2\ell+1) \omega_\ell(\mu^2) P_\ell(0) \left(1 + O(1/J)\right)
\end{multline}
provided the argument of the Legendre polynomial on the second line is sufficiently large and negative. This term dominates over the exponential contribution of the crossing symmetry equation whenever
\begin{equation}
\label{rhoequationgrowslargeJ}
 0 < s - 4 <  \frac{1}{2}(\mu^2 - 4) - t_c 
\end{equation}
where the left inequality simply follows from the need to impose dual positivity only for physical kinematics.

The problem with this asymptotic term is however that the kernel $K(s,\mu^2, - (\mu^2 - 4)/2;0)$ can have a sign flip in the interval \eqref{rhoequationgrowslargeJ}, and if that happens then dual positivity would necessarily be violated on one side of it. Avoiding the sign flip forces us to choose
\begin{equation}
  \mu^2 \leq 12\,.
\end{equation}
We picked $\mu^2 = 12$. To obey the dual positivity constraints at large $J$ for $4 < s < 8 - t_c$ it then suffices to impose%\BvR{Unsure about the sign}
\begin{equation}
  \sum_\ell (2\ell +1)\omega_\ell(\mu^2) P_\ell(0) < 0
\end{equation}
whereas for larger values of $s$ we need to impose \eqref{largeJalphaconstraint}.

\begin{figure*}[t]
    \centering
    \includegraphics[width=\textwidth]{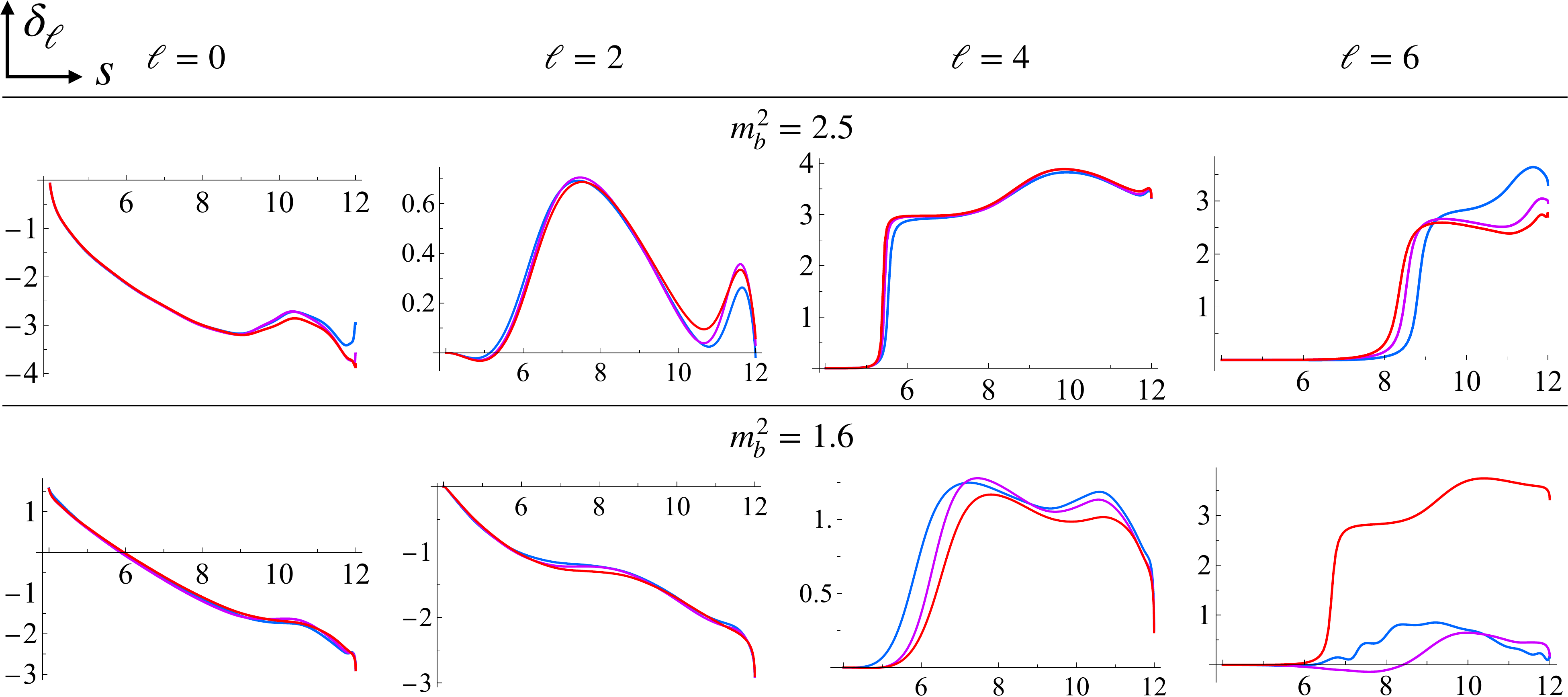}
    \caption{Phase shifts for the amplitude that maximizes the residue at the spin-two bound for two different masses $m_b^2=2.5$, and $m_b^2=1.6$. Different colours correspond to different cutoffs $L_\text{max}$: blue $L_\text{max}=6$, purple $L_\text{max}=8$, and red $L_\text{max}=10$.}
    \label{fig:spin_two_phase_shifts}
\end{figure*}

\subsection{Large energy}
At very large $s$, the two terms in  \eqref{xi_J} have the following behaviour:
\begin{equation}
	\begin{split}
	\sumint\limits_{\ell,v}&\, \omega_\ell(v) R^{(\ell)}_{s \, J} (v) \sim \frac{1}{s^3}\\
	 \sum_{n = 1,3,5, \ldots n_\text{max}}& \alpha_{n} R^{[n]}_{s\, J}(t_c) \sim \frac{1}{s^5}
	\end{split}
\end{equation}
Therefore neglecting the second term, we have the following large energy asymptotics

	\begin{multline}
	\xi_J(s) \sim \frac{2 J + 1}{15 \pi s^3} \int dv \, \omega_0(v) (8-16v +5v^2) \\+ 5 \,\omega_2 (v)(-4+v)^2
	\end{multline}
Thus the large energy constraint is independent of $J$ and it follows from	$\xi_J(s) \leq 0$ that
\begin{equation}
	 \int dv \, \omega_0(v) (8-16v +5v^2) \\+ 5 \,\omega_2 (v)(-4+v)^2 \leq 0.
\end{equation}
While we did not impose these constraints in our numerics, we did check that they were satisfied by the dual solutions that we obtain.

\section{Spin two bound state phenomenology from dual}
\label{sec:phenospin2}

In this Appendix, we study the physics of the extremal dual amplitudes saturating the bound in figure~\ref{fig:spin2_pole} with a single spin 2 bound state.
The three kinks divide the plot in four regions.
In the region $0<m_b^2\lesssim 0.6$, the bound has a strong dependence on the number of Roy equations $L_\text{max}$ imposed. Our conjecture is that in this (unphysical) region it is not possible to have a bound state singularity. The region $m_b^2\gtrsim2$ is the most stable numerically and easier to study. The two regions below $m_b^2\lesssim2$ are separated by a kink around $m_b\simeq 1$. In both regions the phase shifts have a singular threshold behaviour that makes convergence harder.
%$0.6 \lessapprox m_b^2 \lessapprox 1$, $1\lessapprox <m_b^2 \lessapprox 2$
In figure \ref{fig:spin_two_phase_shifts} we plot the phase shifts for $\ell=0,\dots,6$ of two typical amplitudes in the two regions, for $m_b^2=2.5$, and $m_b^2=1.6$.

\subsection{Extremal amplitudes with $m_b^2\gtrsim2$}
\begin{figure}[t]
    \centering
    \includegraphics[width=0.8\linewidth]{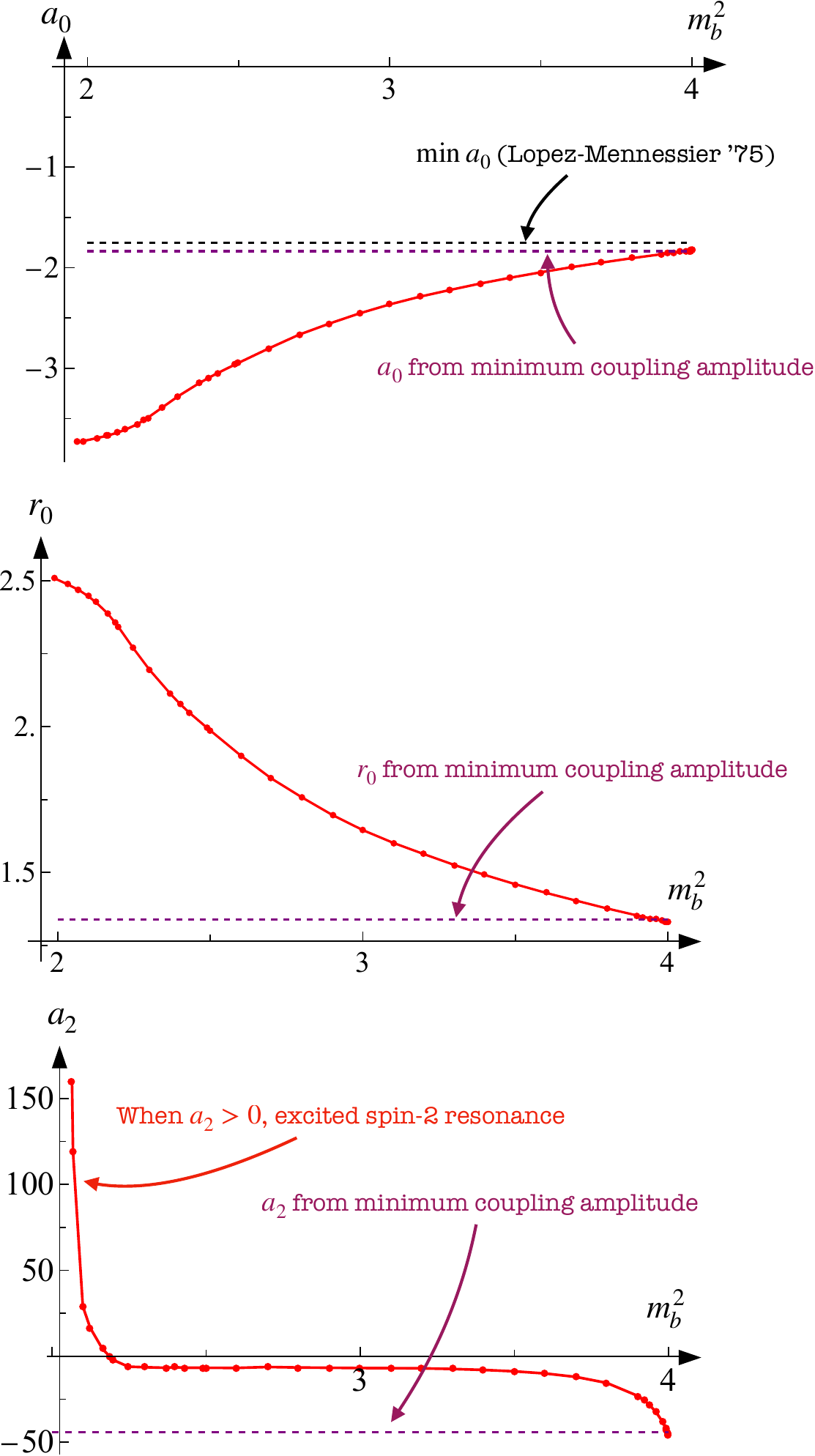}
    \caption{Threshold parameters of the extremal amplitudes saturating the maximum spin two residue as a function of the position of the pole $m_b^2>2$.}
    \label{fig:scatt_eff_range_spin_two_s_wave}
\end{figure}

We start with the easy region $m_b^2\gtrsim2$, for which the top row in figure \ref{fig:spin_two_phase_shifts} is a representative example.
For $\ell=0$ the phase shifts start at threshold as $\delta_0(k_s)\sim k_s$, whereas for $\ell=2$ the threshold expansion is well described by $\delta_2(k_s)\sim k_s^3$.
The corresponding threshold expansion of the lowest spin partial amplitudes takes the form
\begin{eqnarray}
&&f_0(k_s)=f_0^{(0)}+\frac{i}{2}(f_0^{(0)})^2 k_s+f_0^{(2)}k_s^2+\nonumber\\
&&+ \frac{i}{8}f_0^{(0)}((f_0^{(0)})^3-2 f_0^{(0)}+8 f_0^{(2)})k_s^3+f_0^{(4)}k_s^4+\dots\nonumber\\
\label{eq:spin-zero-expansion}
\end{eqnarray}
\begin{eqnarray}
&&f_2(k_s)=k_s^4\left(f_2^{(0)}+f_2^{(2)}k_s^2+f_2^{(2)}k_s^4+i \frac{(f_2^{(0)})^2}{2}k_s^5+\dots\right)\nonumber\\
\label{eq:spin-two-expansion}
\end{eqnarray}
Generalizing definition in eq.~\eqref{eq:definition_scatt_length} to 
\begin{equation}
k_s^{2\ell+1}\cot\delta_\ell(k_s)=\frac{1}{a_\ell}+\frac{1}{2}r_\ell k_s^2+\dots
\end{equation}
and using the fact that $e^{2i\delta_\ell(k_s)}=1+i\tfrac{k_s}{\sqrt{1+k_s^2}}f_\ell(k_s)$, we obtain for the $\ell=0$ wave
\begin{eqnarray}
a_0&=&\frac{f_0^{(0)}}{2},\nonumber\\
r_0&=&\frac{2-(f_0^{(0)})^2}{f_0^{(0)}}-4\frac{f_0^{(2)}}{(f_0^{(0)})^2},
\end{eqnarray}
and similarly for $\ell=2$.
Fitting the phase shifts with the ansatz~\eqref{eq:spin-zero-expansion} and~\eqref{eq:spin-two-expansion} we can extract scattering length and effective range parameters as a function of the spin-two bound state position of the pole.
The results are in figure~\ref{fig:scatt_eff_range_spin_two_s_wave}. 

Interestingly, we find that as $m_b^2\to 4$, both scattering lengths $a_0$ and $a_2$ as well as the effective range $r_0$ tend to the values extracted from the amplitude that minimizes the quartic coupling in absence of poles (studied for instance in~\cite{Bonnier:1975jz,Paper3,Guerrieri:2021tak,EliasMiro:2022xaa}).
The quartic coupling is defined as $32\pi \lambda=T(s=t=u=4/3)$, and in the absence of poles it must take values in the interval~\cite{Guerrieri:2021tak}
\begin{equation}
-8.1<\lambda<2.72.
\end{equation}
From a primal perspective, it is hard to make the optimization problem converge, see~\cite{EliasMiro:2022xaa} for a recent attempt to improve the convergence.
The triple coincidence in figure~\ref{fig:scatt_eff_range_spin_two_s_wave} leads us to the conjecture that the theory saturating the minimum bound on the quartic coupling is obtained by continuing the spin-two pole up to the threshold $s=4m^2$. It would be interesting to develop a primal ansatz featuring a spin-two pole singularity following the attempt in~\cite{Auberson:1979ye}.

%Threshold pole singularities are not an exotic feature of this amplitude. They appear in the 

The absolute minimum on the $s$-wave scattering length found in~\cite{Lopez:1975ca} is slightly stronger $a_0\geq -1.75$ than the one extracted from the dual amplitude minimizing the quartic coupling, which is $a_0\simeq -1.84$. A natural conjecture would have been that the theory with most negative quartic coupling is also the theory with the most negative threshold behaviour. It might be interesting to look more in detail at this discrepancy and find an explanation for it. Moreover, to the best of our knowledge there exist no rigorous bound on the effective range. Here we estimate $r_0\simeq 1.3$ in absence of bound state poles. By analogy, we might guess that this value is also close to being extremal. We have a similar situation for the $a_2$ scattering length.

Scattering lengths for higher spins $a_{\ell\geq 4}$ are positive, and the corresponding phase shifts show the typical resonant behaviour jumping by $\pi$ around the resonance position, see figure~\ref{fig:spin_two_phase_shifts}. In the narrow-width approximation the resonance mass can be found at the point where the phase passes through $\pi/2$. For simplicity, here we extract approximately the resonance by solving the equation $\delta_\ell(m^2)=\pi/2$ for all points where the phase grows enough. The results are summarized in figure~\ref{fig:Regge_plot_spin2pole} in the main text. 

\begin{figure}[t]
    \centering
    \includegraphics[width=\linewidth]{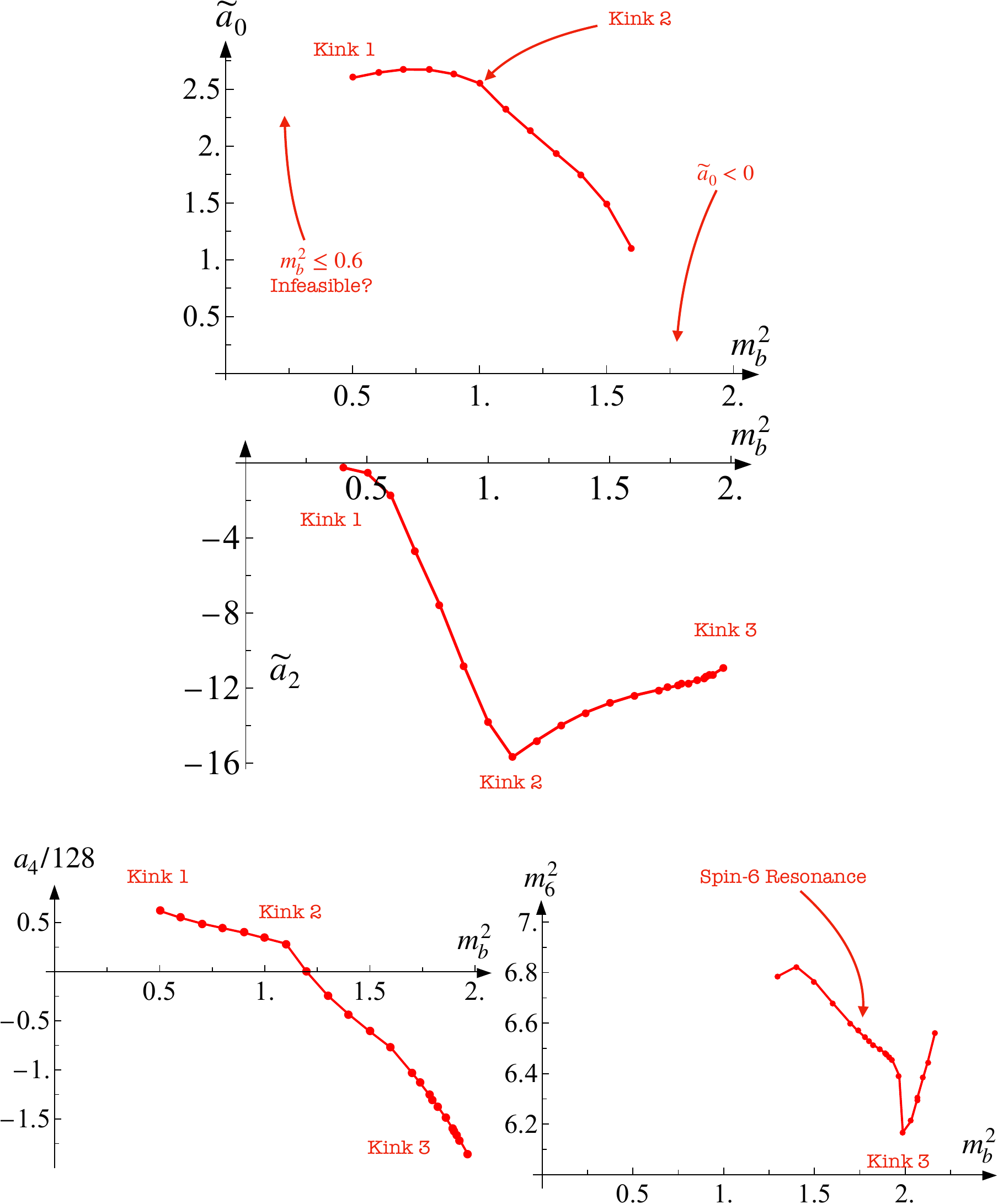}
    \caption{Threshold parameters of the extremal amplitudes saturating the maximum spin two residue as a function of the position of the pole $m_b^2<2$.}
    \label{fig:threshold_params_at_spin2_bound}
\end{figure}

\subsection{Extremal amplitudes with $m_b^2\lesssim2$}
\label{sec:extreamal_less_than_two}

\begin{figure*}[t]
    \centering
    \includegraphics[width=\textwidth]{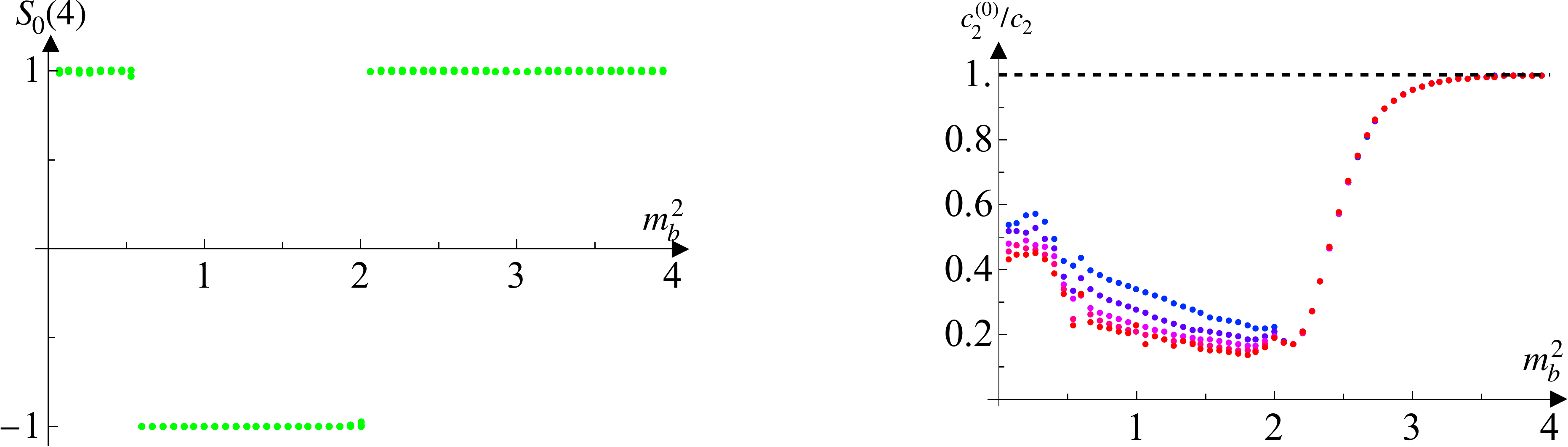}
    \caption{On the left, the value of $S_0(4)$ as a function of the position of the scalar pole $m_b^2$. On the right, the ratio $c_2^{(0)}/c_2$ as a function of $m_b^2$. Different colours correspond to different values of $N_\text{max}$, ranging from 7 (blue) to 11 (red).}
    \label{fig:spin_0_physics}
\end{figure*}

When $m_b^2\lesssim2$, the amplitudes that maximize the residue at the pole have a singular threshold behaviour. This is evident in the $\ell=0$ wave where the $\delta_0$ phase shifts start at $\pi/2$ as in the example in figure~\ref{fig:scatt_eff_range_spin_two_s_wave} for $m_b^2=1.6$.
The most general threshold expansion of the partial amplitude compatible with unitarity takes the form
\begin{eqnarray}
&&f_0(k_s)=\frac{2i}{k_s}+f_0^{(0)}+\frac{i}{2}(2-(f_0^{(0)})^2)k_s+f_0^{(2)}k_s^2+\nonumber\\
&&-\frac{i}{8}((f_0^{(0)})^4{-}2 (f_0^{(0)})^2{+}2{-}32 f_0^{(0)} f_0^{(2)})k_s^3+f_0^{(4)} k_s^4+\dots\nonumber\\
\end{eqnarray}
We can formally define a scattering length and an effective range using the expansion
\begin{equation}
\frac{1}{k_s}\cot\delta_0(k_s)=\frac{1}{\widetilde a_0}+\frac{1}{2}\widetilde r_0 k_s^2+\dots
\end{equation}

Less clear from the figure is the threshold behaviour of the $\delta_2$ wave.
We have numerically found that the best fit is given by an ansatz of the form
\begin{equation}
f_2(k_s)=k_s^2\left(f_2^{(0)}+f_2^{(2)}k_s^2+\frac{i}{2}(f_2^{(0)})^2 k_s^3+f_2^{(4)}k_s^4+\dots\right).
\label{eq:f2}
\end{equation}
The above expression is not arbitrary. It is the most general expansion compatible with elastic unitarity and the least singular behaviour compared to the regular expansion~\eqref{eq:spin-two-expansion}.

In figure~\ref{fig:threshold_params_at_spin2_bound}, we summarize the information extracted from fitting the singular threshold expansions in this region.
We extract the threshold parameters $\widetilde a_0$ and $\widetilde a_2$ in the mass range $0.6\leq m_b^2\leq 2$. 
We find that the second kink in figure~\ref{fig:spin2_pole} is indeed correlated to a change in behaviour of $\widetilde a_2$ where it attains its minimum value.

Higher spins have a regular threshold expansion. In the same figure~\ref{fig:threshold_params_at_spin2_bound} we also plot the spin four $a_4$ scattering length. 
In between the first and the second kink it is positive, then rapidly becomes negative. Although the phase shift becomes large, we do not observe the typical resonant behaviour.
Starting with $\ell=6$ we do observe a resonance, and we plot its mass as a function of $m_b^2$.

\section{Scalar bound state phenomenology from primal}
\label{scalar_pheno}

In this Appendix we study in detail the physics of the extremal amplitudes that maximize the residue of the single scalar bound state extracted using the primal S-matrix Bootstrap in figure~\ref{fig:scalar_pole}.

The primal amplitude is parameterized using the wavelet ansatz introduced in~\cite{EliasMiro:2022xaa}.
%with approximately $N_\text{max}^3/7$ terms.  
Such an ansatz trivially satisfies crossing and maximal analyticity, but not unitarity. We impose unitarity by projecting the ansatz in partial waves and demanding that $|S_\ell|^2 \leq 1$ for any $\ell$ and for any $s>4$. In practice, we impose unitarity up to a maximal spin $L_\text{max}$ and on a finite grid in $s$. To control the higher spin tail $\ell>L_\text{max}$, we also impose fixed $t$ positivity constraints of the form $\Im T(s,0<t<4)\geq 0$ for $s>4$.

\begin{figure*}[t]
    \centering
    \includegraphics[width=\textwidth]{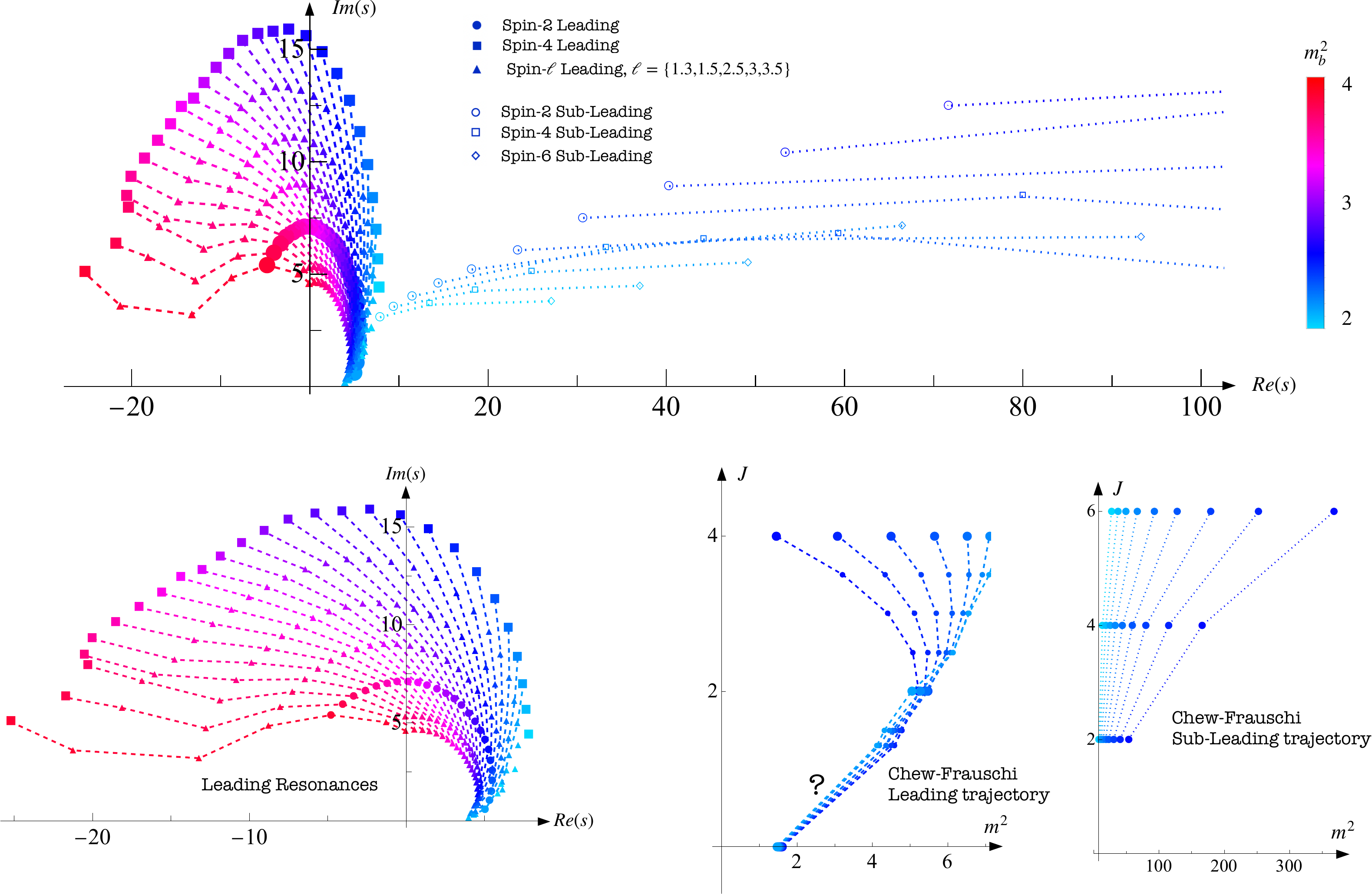}
    \caption{Spectrum of resonances in the complex $s$-plane of the extremal amplitudes as a function of the mass of the scalar bound sate $m_b^2$. Each colour correspond to a different amplitude. We connect with lines the resonances we believe belong to the same trajectory: dashed for what we call `leading', dotted for the `sub-leading'. Physical spins are indicated by larger markers. In the two bottom right panels we show the Chew-Frautschi plots for the leading and sub-leading trajectories. For the leading one we have tentatively extrapolated up to spin zero. It would be interesting to see whether this is correct, as we do not have sufficient resolution with the current numerics.}
    \label{fig:peacock_plot}
\end{figure*}

The primal ansatz also contains \emph{threshold singularity} terms of the form 
\begin{equation}
T(s,t) \supset \alpha_{th}\left(\frac{1}{\rho_{4/3}(s){-}1}+\frac{1}{\rho_{4/3}(t){-}1}+\frac{1}{\rho_{4/3}(u){-}1}\right),
\end{equation}
where 
\begin{equation}
\rho_{s_0}(s)=\frac{\sqrt{4-s_0}-\sqrt{4-s}}{\sqrt{4-s_0}+\sqrt{4-s}}.
\end{equation}
These terms are permitted by unitarity, and the allowed values for the residue $-64 \pi\sqrt{3/2}\leq \alpha_{th}\leq 0$ are consequence of unitarity at threshold.

In the plot in figure~\ref{fig:scalar_pole} we can identify two special points. First, we have the maximum allowed residue for a scalar bound state which is attained at $m_b^2=2$. 
This situation is reminiscent of what happens in $1+1$ dimensions: in that case, the maximum allowed value is also attained at $m_b^2=2$, but it is infinite because of the exact cancellation between the $s$ and $t$-channel poles both sitting at $m_b^2=4-m_b^2=2$ with opposite residues. In $3+1$ dimensions the $t$-channel pole is smeared into a log-singularity after partial wave projection, preventing exact cancellation. Second, we have a kink at $m_b^2\simeq 0.6$, at approximately the same position as in the spin-two bound state plot~\ref{fig:spin2_pole}. As in that case, we believe below this point it is impossible to have a bound state pole with finite residue. Our belief is supported by the fact that the primal bound has a strong spin cutoff dependence.

In what follows we will correlate the kinks in the residue plot with features of the extremal amplitudes that will help us to shed a light on their physical content. In figure~\ref{fig:spin_0_physics}, we start a first characterization.
In the plot on the left, we have the value of $S_0(4)$ as a function of $m_b^2$. It saturates alternatively the unitarity inequality with two jumps that happen at the two cusps in figure~\ref{fig:scalar_pole}. The threshold singularity determines the value of the spin-zero S-matrix at $s=4$: $S_0(4)=1$ when $\alpha_{th}=0$, while $S_0(4)=-1$ when $\alpha_{th}=-64 \pi\sqrt{3/2}$.

We can make a close analogy between the behaviour of the spin-zero S-matrix in 3+1 dimensions and the CDD pole factors in 1+1 dimensions. CDD pole factors saturate the bound on the maximum residue of a bound state in 1+1 dimensions. Their analytic expression is
\begin{equation}
S_{CDD}=(-1)^{\theta(2-m_b^2)}\frac{\sqrt{m_b^2(4-m_b^2)}+\sqrt{s(4-s)}}{\sqrt{m_b^2(4-m_b^2)}-\sqrt{s(4-s)}},
\end{equation}
where the sign is fixed such that the residue of the $s$-channel pole is positive. Indeed, this implies that $S_{CDD}(4)=1$ when $m_b^2>2$, and $S_{CDD}(4)=-1$ for $m_b^2<2$. In $3+1$ dimensions, the mechanism to obtain a negative spin-zero S-matrix at threshold is to introduce a threshold singularity.

\begin{figure*}[t]
    \centering
    \includegraphics[width=\textwidth]{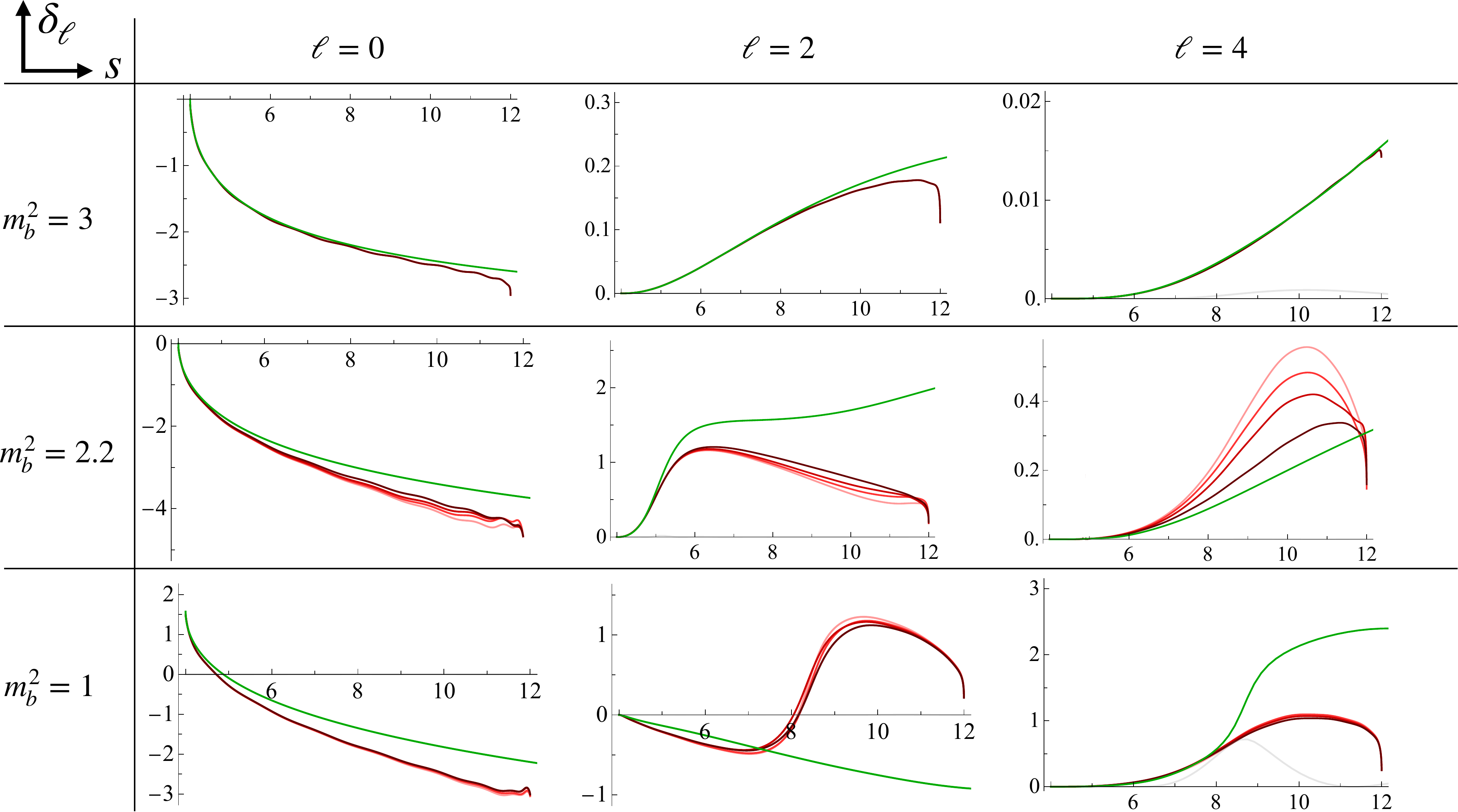}
    \caption{Phase shifts of the extremal amplitudes maximizing the residue at the scalar bound state pole extracted from primal (in green) and dual (in red). We plot in red shades different values of $L_\text{max}$ ranging from 4 to 10. The light grey line is the inelasticity $1-|S_\ell(s)|^2$ extracted from primal. When $1-|S_\ell(s)|^2=0$ means that primal phase shifts are well converged.}
    \label{fig:primal_dual_scalar_deltas}
\end{figure*}

To check the importance of higher spins to the physics of the extremal amplitudes it is useful to study the spin-zero dominance. 
For instance, we consider the following dispersive integral
\begin{equation}
c_2= \sum_{\ell=0}^\infty c_2^{(\ell)}=\int_4^\infty \frac{ds}{s^3}\sum_{\ell=0}^\infty 16\pi(2\ell+1)\Im f_\ell(s)\geq 0.
\end{equation}
The parameter is positive and can be interpreted as an integral of the total cross section $\Im T(s,0)/s^3\equiv \sigma_\text{tot}/s^2$.
It can also be related to the second derivative of the amplitude $c_2\equiv \partial^2_s T(s,0)|_{s=0}$. 
In figure~\ref{fig:spin_0_physics}, on the right, we plot the ratio $c_2^{(0)}/c_2$, where $c_2^{(\ell)}$ is defined to be the spin-$\ell$ contribution to the above sum rule.
The different colours correspond to different $N_\text{max}$, ranging from $N_\text{max}=7$ (with 94 free parameters in the ansatz) in blue to $N_\text{max}=11$ (286 free parameters) in red.
In the range $2\leq m_b^2<4$ this ratio is smooth and well converged in $N_\text{max}$. As the bound state moves from the threshold to the cusp at $m_b^2=2$, this ratio decreases, signaling that the higher spins become important and eventually dominate the cross section. For $m_b^2<2$ convergence in $N_\text{max}$ is significantly slower, and the higher spins dominate. 
This slower convergence seems related to an important feature of the extremal amplitudes with $m_b^2<2$: the presence of a threshold singularity in the spin-two. This behaviour is similar to the one encountered in Appendix~\ref{sec:extreamal_less_than_two} where the spin-two amplitude behaves at threshold as $f_2(s)\sim \mathcal{O}(s-4)$. We believe to have better convergence it is crucial to write a primal ansatz admitting such behaviour.

To better understand the physics of higher spins, we should look at the spectrum of resonances. In figure~\ref{fig:peacock_plot} we plot the spectrum of resonances in the complex $s$-plane. We denote resonances of different spins and belonging to different trajectories with different plot-markers. However, points with the same colour belong to the same amplitude. We also connect with dashed and dotted lines the resonances belonging to the same trajectory. We focus on the region $m_b^2>2$.

We observe two different kind of trajectories. The one we call `Leading' contains resonances that, as we move $m_b^2$ from 2 to 4, have a different physical interpretation.
When $m_b^2\sim 2$, they can be described as weakly coupled light resonances. When $m_b^2\sim 3$, they acquire a large imaginary part and become highly unstable. As $m_b^2\to 4$, they approach the left-cut region, and we are not sure of their interpretation. To better follow these highly curved trajectories we use the Froissart-Gribov representation and continue our ansatz to complex spins using the formula
\begin{equation}
f_\ell(s)=\frac{1}{32\pi}\int_4^\infty dt \frac{8}{\pi(s-4)}Q_\ell\left(1+\frac{2t}{s-4}\right)T_t(s,t),
\end{equation}
valid for $\Re\ell>0$. While it is difficult to prove it, numerical results do suggest that the spin-zero bound state poles also belong to the various trajectories.

Beyond the leading, we also observe nearly linear `Sub-Leading' trajectories. Resonances in this case nicely align as shown in the Chew-Frautschi plot in the bottom right panel of figure~\ref{fig:peacock_plot}.

\subsection{Comparing primal and dual phase shifts}

In this section we compare in detail primal and dual phase shifts in the interval $4<s<12$.
In figure~\ref{fig:spin_two_phase_shifts} we plot primal (in green) and dual (in red) phase shifts for $\ell=0,2,4$ at three benchmark points. The points are chosen as representatives of the different regions along the boundary in figure~\ref{fig:scalar_pole}. We immediately observe a correlation between the duality gap and the discrepancy between the phase shifts obtained using the two methods. In the region $2 \ll m_b^2<4$, there is a quantitative agreement of the phase shifts up to the end-point of the window $s=12$ where the dual problem stops converging.
In the region close to $m_b^2=2$, but slightly above, the dual phase shifts depend strongly on the number of Roy equations $L_\text{max}$, and the gap between primal and dual phase shifts is non-negligible. Despite the gap, however, qualitatively the phase shifts have the same behaviour, and as we increase $L_\text{max}$ they seem to better agree. 
For $m_b^2<2$ the gap is indeed large, up to 20\% of the total bound. This is somehow expected due to the high-spin dominance shown in figure~\ref{fig:spin_0_physics}. Phase shifts, except for the spin zero wave look quite different. 

\begin{figure}[t]
    \centering
    \includegraphics[width=0.9\linewidth]{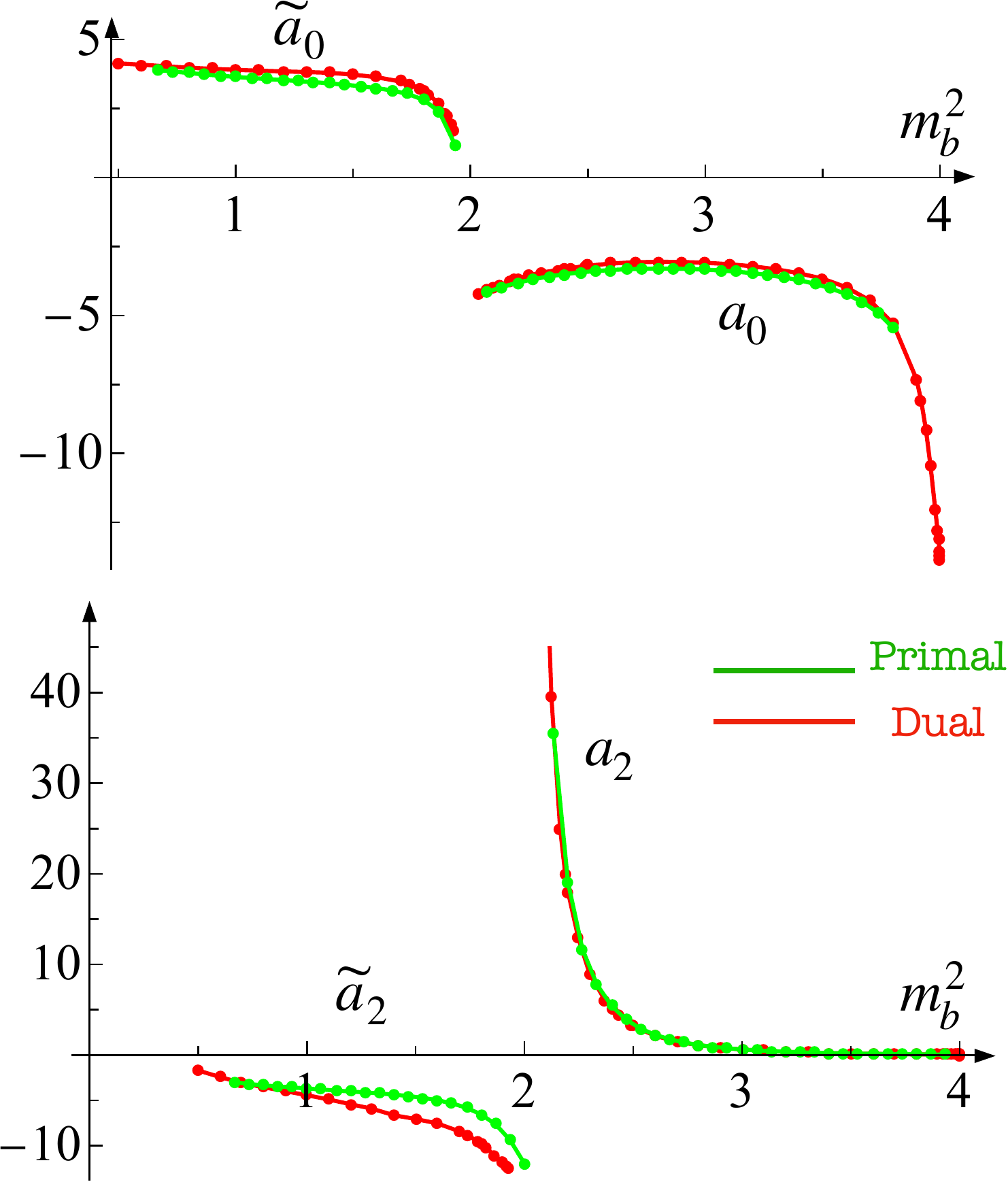}
    \caption{Threshold parameters extracted from the primal (in green) for $N_\text{max}=11$, and from the dual (in red) for $L_\text{max}=10$.}
    \label{fig:primal_dual_threshold_comparison}
\end{figure}

We conclude this section by comparing the primal and dual estimate of the threshold coefficients. By looking at figure \ref{fig:spin_0_physics}, we can already assume that they will agree better. In figure~\ref{fig:primal_dual_threshold_comparison} we check this is the case for both $\delta_0$, and $\delta_2$ by extracting respectively in the region $m_b^2>2$ the scattering lengths $a_0$ and $a_2$, and in the region $m_b^2<2$, the anomalous scattering lengths $\widetilde a_0$ and $\widetilde a_2$.
Threshold parameters for the $\ell=0$ wave agree in both regions. The biggest discrepancy comes from the $\ell=2$ parameters and in the region $m_b^2<2$. It would be interesting to understand this discrepancy and close the gap. From the primal side it would be important to design an ansatz with a threshold behaviour general enough to account for a singular phase shift as the one in~\ref{eq:f2}. From the dual side, it would be crucial to extend the integration domain of the dispersion relations and access to a larger region in $s$ to better describe the higher spin physics.

\bibliographystyle{apsrev4-1}
\bibliography{dual_bootstrap}

\end{document}